\documentclass{aa}

\def \dj{d\kern-0.4em\char"16\kern-0.1em}
\def \Dj{\mbox{\raise0.3ex\hbox{-}\kern-0.4em D}}
\newcommand{\ind}[2]{#1_\mathrm{#2}}
\newcommand{\Rblr}[2]{\ind{R}{#1}^{\mathrm{#2}}}
\usepackage{graphicx}
\usepackage{epstopdf}
\usepackage{txfonts}
\usepackage{siunitx}
\usepackage{multirow}
\usepackage{amsfonts}
\usepackage{bm}
\usepackage{hyperref}

\usepackage{etoolbox}
\makeatletter
\patchcmd\@combinedblfloats{\box\@outputbox}{\unvbox\@outputbox}{}{%
  \errmessage{\noexpand\@combinedblfloats could not be patched}%
}%
\makeatother

\DeclareSIUnit\solarmass{\ensuremath{M_\sun}}
\DeclareSIUnit\angstrom{\ensuremath{\AA}}
\begin{document}

   \title{AGN Black Hole Mass Estimates Using Polarization in Broad Emission Lines}
   \author{\Dj.\@ Savi\' c \inst{1,2}
          \and
          R.\@ Goosmann\inst{1}
          \and
          L.\@ \v C.\@ Popovi\' c\inst{2,3}
          \and
          F.\@ Marin\inst{1}
          \and
          V.\@ L.\@ Afanasiev\inst{4}
          }

  \institute{Observatoire Astronomique de Strasbourg, Universit\'e de Strasbourg, CNRS, UMR 7550, 11 rue de l’Universit\'e, 67000 Strasbourg, France
	  \and
          Astronomical Observatory Belgrade, Volgina 7, 11060 Belgrade, Serbia
          \and
          Department of Astronomy, Faculty of Mathematics, University of Belgrade, Studentski trg 16, 11000 Belgrade, Serbia
          \and
          Astrophysical Observatory of the Russian Academy of Sciences, Nizhnij Arkhyz, Karachaevo-Cherkesia 369167, Russia
          \\
          \\
          \email{djsavic@aob.rs}
          \\
          Received 06/11/2017; accepted 09/01/2018
          }
  \date{}

   \abstract{The innermost regions in active galactic nuclei (AGNs) were not being spatially resolved so far but spectropolarimetry can provide us insight about their hidden physics and the geometry. From  spectropolarimetric observations in broad emission lines and assuming equatorial scattering as a dominant polarization mechanism, it is possible to estimate the mass of supermassive black holes (SMBHs) residing in the AGN center.}
        {We explore the possibilities and limits and put constraints on the usage of the method for determining SMBH masses using polarization in broad emission lines by providing more in-depth theoretical modeling.}
        {We use the Monte Carlo radiative transfer code \textsc{stokes} for exploring polarization properties of Type 1 AGNs. We model equatorial scattering using flared-disk geometry for a set of different SMBH masses assuming Thomson scattering. In addition to the Keplerian motion which is assumed to be dominant in the broad line region (BLR), we also consider cases of additional radial inflows and  vertical outflows.}
        {We model the profiles of polarization plane position angle $\varphi$, degree of polarization and total unpolarized line for different BLR geometries and different SMBH masses. Our modeling confirms that the method can be widely used for Type-1 AGNs when viewing inclinations are between $25^\circ$ and $45^\circ$. We show that the distance between the BLR and scattering region (SR) has a  significant impact on the mass estimates and the best mass estimates are when the SR is situated at the distance 1.5--2.5 times larger than the outer BLR radius.}
        {Our models show that if Keplerian motion can be traced through the polarized line profile, then the direct estimation of the mass of the SMBH can be performed. When radial inflows or vertical outflows are present in the BLR, this method can still be applied if velocities of the inflow/outflow are less than \SI{500}{\kilo\meter\per\second}. We also find that models for NGC 4051, NGC 4151, 3C 273 and PG0844+349 are in good agreements with observations.}


   \maketitle
%

\section{Introduction}
Active galactic nuclei (AGNs) are known to be one of the most powerful and steady radiation sources in the Universe. The huge amount of energy is produced by the accretion of matter onto supermassive black holes \citep[SMBHs,][]{1969Natur.223..690L} which mass ranges from \SI{d6}-\SI{d10}{\solarmass} \citep{1995ARA&A..33..581K}. The energy released by the growth of the black hole exceeds the binding energy of the host galaxy bulge \citep{2012ARA&A..50..455F}. Thus, we can expect that AGNs have a strong feedback on it's environment due to the strong interaction between energy and radiation produced by accretion and the surrounding gas of the host galaxy. This can lead to heating or ejection of the interstellar gas, which can prematurely terminate star formation in the galaxy bulge. This is well supported by the observed correlation between the mass of the central SMBH with luminosity, stellar velocity dispersion $\sigma_*$ or bulge mass \citep{2013ARA&A..51..511K}, which indicates that the coevolution of the SMBHs and the host galaxies \citep{2014ARA&A..52..589H} exists. Measuring SMBH masses is crucial task in order to understand how are they linked with the evolution of galaxies and AGNs.

Our ability to estimate the masses of AGN central black holes has significantly advanced in the recent years \citep[see e.\,g.][for a review]{2014SSRv..183..253P}. Several methods (both direct and indirect) have been developed. Direct methods are those for which mass of the black hole is obtained from stellar dynamics by studying the motions of individual stars around the black hole \citep{2010RvMP...82.3121G,2012Sci...338...84M}, or gas dynamics \citep[see e.\,g.][]{1995Natur.373..127M}. Indirect methods use observables that are tightly correlated with black hole mass. One such example is $\ind{M}{bh} - \sigma_*$ relation \citep{2000ApJ...539L..13G,2001ApJ...555L..79F,2005SSRv..116..523F}. The most reliable (direct) mass measurements of SMBHs come from reverberation mapping \citep{1982ApJ...255..419B} of broad emission lines in about sixty AGNs \citep{2015PASP..127...67B}. However, reverberation mapping includes some unknown assumptions of the Keplerian motion and the photoionization as dominant physical process in the BLR.

A method of the AGN black hole mass estimation using polarization in the broad lines given by \citet[][hereafter AP15]{2015ApJ...800L..35A}, assumes that broad line photons are emitted from the disk-like region undergoing a Keplerian motion, after which are being scattered by the surrounding dusty torus, resulting in polarization in the broad emission lines. This method is in a good agreement with reverberation one and offers a number of advantages over traditional reverberation mapping. This method needs only one epoch of observations and it is not telescope time consuming as reverberation mapping method. It can be applied to lines from different spectral ranges and thus allowing black hole mass measurements for AGNs at different cosmological epochs \citep[see for more details in][]{2015ApJ...800L..35A}. Note here that in this method, the approximation of one scattering event per line photon was used and the contribution of multiple scattering events were not taken into account. Due to the fact that the polarization is very sensitive to kinematics and geometrical setup \citep{2007A&A...465..129G}, the full treatment of 3D radiative transfer with polarization is required to test this method. The aim of this work is to explore the AP15 method applying more accurate radiative transfer modeling. First we modeled the polarization in the broad lines using the \textsc{stokes} code, and after that we compare the calculated polarization with observed in four Type 1 AGNs.

The paper is organized as followed: in Section \ref{S.method} we give the description of the method for mass determination using polarization in broad lines, in Section \ref{S.model} we describe parameters used for models, in Section \ref{S.observations} we give basic informations of the observed objects used here; our results are given in Section \ref{S.results}. In Section 6 we discuss the obtained results and in Section 7 we shortly outline main conclusions.

\section{AP15 method} \label{S.method}
According to the unified model \citep{1993ARA&A..31..473A,1995PASP..107..803U}, every AGN hosts an accreting SMBH surrounded by a dusty torus along the equatorial plane. When the line of sight towards the central engine is unobscured, in Type-1 objects, permitted broad spectral lines are prominent in the optical spectra. Broad lines are emitted from the broad-line region (BLR) -- high density clouds ($\sim\SI{d10}{\centi\meter\tothe{-3}}$) situated around an accreting black hole with a global covering factor of order 0.1 \citep{2013peag.book.....N}. We can expect near-Keplerian motion of the emitting gas in the BLR \citep{2009NewAR..53..140G}. Farther away, the central region is surrounded by geometrically thick toroidal structure of gas and dust with large radial optical depth \citep{1988ApJ...329..702K}. The inner side of the torus is directly illuminated and one can expect an abundance of free electrons in this part.

\begin{figure*}
   \centering
   \includegraphics[width=\hsize]{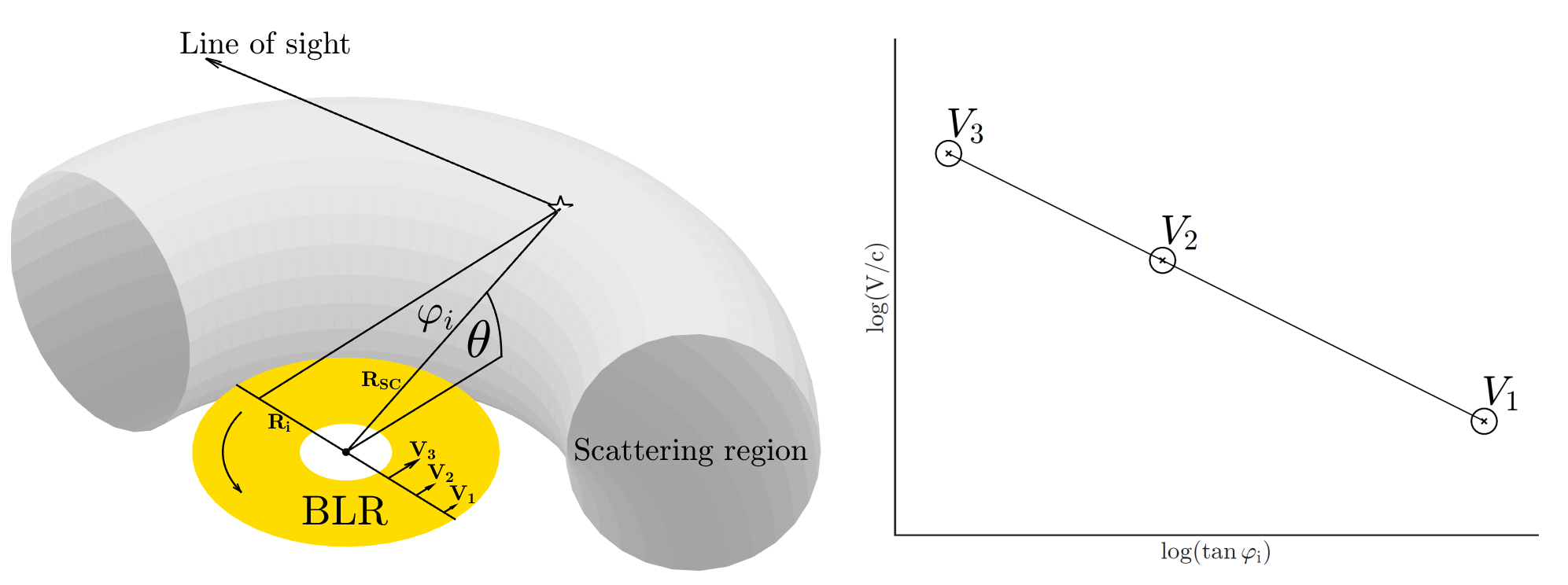}
      \caption{Schematic view of light being scattered from the inner part of the torus (left). Expected relation between $\varphi$ and velocity intensity (right).}
         \label{method}
\end{figure*}

The BLR is surrounded by a co-planar scattering region (SR) which produces polarized broad lines and a characteristic change of polarization plane position angle $\varphi$ across the line profile can be expected \citep{2005MNRAS.359..846S}. According to AP15, $\varphi$ in the broad emission line is affected by the velocity field in the BLR and have specific linear relationship between $\log V$ and $\log(\tan\\varphi)$. As shown in Fig.\@ \ref{method} (right), if we have separation in the velocity field, it will affect $\varphi$ across the line.

\noindent Let us shortly recall the method. If the motion in the BLR is Keplerian, for the projected velocity in the plane of the scattering region, we can write \citep[see][AP15]{2014MNRAS.440..519A}:
\begin{equation}
\label{V_i}
V_i = V_i^{\mathrm{rot}}\cos{\theta} = \sqrt{\frac{G\ind{M}{BH}}{R_i}}\cos{\theta}
\end{equation}

\noindent where $V_i$ is the rotational velocity of emitting gas, $\ind{M}{BH}$ is the BH mass and $R_i$ is the distance from the center of the disk; $G$ is the gravitational constant and $\theta$ is the angle between the disc and the plane of scattering \citep{2015ApJ...800L..35A}. $R_i$ can be connected with the corresponding polarization angle $\varphi_i$ as $R_i =\ind{R}{sc}\tan{\varphi_i}$, where $\ind{R}{sc}$ is the distance from the center to the SR (Fig.\@ \ref{method}). When we substitute this into Eq.\,(\ref{V_i}) taking into account contribution of the different parts of the disk, the velocity-angle dependence can be transformed in the following way:

\begin{equation}
 \log{\frac{V_i}{c} = a - 0.5\log{\left(\tan{\left(\varphi_i\right)}\right)}},
\end{equation}
\noindent where $c$ is the speed of light. The expected relation between velocity and $\varphi$ is shown in Fig.\@ \ref{method} (right). The constant $a$ is related to the black hole mass as

\begin{equation}
 a = 0.5\log\frac{G\ind{M}{BH}\cos^2{\theta}}{c^2\ind{R}{sc}}.
\end{equation}

In the case of a thin SR (equatorial scattering region), a good approximation would be to take $\theta\sim 0$. In this case, the relation between velocities and $\varphi$ does not depend on the inclination since the BLR is emitting nearly edge-on oriented line light to the SR. From the previous equation, the mass of the black hole can be calculated as

\begin{equation}
 \label{eq:Mbh}
 \ind{M}{BH} = 1.78 \times 10^{2a + 10}\frac{\ind{R}{sc}}{\cos^2{\theta}}\SI{}{\solarmass} \approx 1.78 \times 10^{2a + 10}\ind{R}{sc}\, \left[\SI{}{\solarmass}\right],
\end{equation}

\noindent or

\begin{equation}
 \log\frac{\ind{M}{BH}}{\SI{}{\solarmass}} = \left(10 + 2a\right)\log\left(1.78\ind{R}{sc}\right),
\end{equation}

\noindent where $\ind{R}{sc}$ is in light days.

\section{Simulation of equatorial scattering} \label{S.model}
\subsection{Radiative transfer code}
We have used the radiative transfer code \textsc{stokes} \citep{2007A&A...465..129G,2012A&A...548A.121M,2015A&A...577A..66M} for investigation of polarization in the broad emission line in AGNs. It is a 3D radiative transfer code based on Monte Carlo approach. It follows single photons from their creation inside the emitting region through processes such as electron or dust scattering until they become absorbed or until they manage to reach distant observer. Initially, it was developed to study the ultraviolet (UV) and optical continuum polarization induced by electron and dust scattering in the radio-quiet AGNs, but it is suitable for studying many astrophysical objects of various geometries \citep{2014sf2a.conf..103M}. We used the latest 1.2 version of the code \textsc{stokes} which is publicly available\footnote{http://www.stokes-program.info/}.

\subsection{Parameters of the model}
In our model, a point-like continuum source is situated in the center, emitting isotropic unpolarized radiation for which the flux is given by a power-law spectrum $F_C \propto \nu^{-\alpha}$ with $\alpha = 2$. Since we are investigating spectral range around specific line, the chosen value for the spectal index $\alpha = 2$ will not affect our research.

The continuum source is surrounded by a BLR which is finally surrounded by a SR. BLR and SR are modeled using flared-disk geometry with half-opening angle from the equatorial plane of $15^\circ$ (covering factor $\sim 0.1$) for the BLR and $35^\circ$ for the SR. High covering factor for the SR is necessary in order to obtain the observed profile of the $\varphi$. Low covering factor of the SR gives very small amplitude in the $\varphi$ profile. For the BLR inner radius $\Rblr{in}{BLR}$, we adopted the value obtained by reverberation mapping \citep{2005ApJ...629...61K,2006ApJ...651..775B,2013ApJ...767..149B}. The BLR outer radius was set by dust sublimation (predicted by \citealt{1993ApJ...404L..51N}):

\begin{equation}
\Rblr{out}{BLR} = 0.2\ind{L}{bol,46}^{0.5},
\end{equation}

\noindent where $\ind{L}{bol,46}$ is bolometric luminosity given in \SI{e46}{ergs \second\tothe{-1}}. Bolometric luminosity is approximated from optical nuclear luminosity \citep{2012MNRAS.427.1800R}:

\begin{equation}
  \log{\ind{L}{iso}} = 4.89 + 0.91\log{\ind{L}{5100}},
\end{equation}
\noindent where $\log{\ind{L}{5100}}$ is the optical nuclear luminosity. After correcting for average viewing angle, we obtain $\log{\ind{L}{bol}} = 0.75\log{\ind{L}{iso}}$. In our model, the BLR is transparent to photons i.e., photons can freely travel from the inner side to the outer side of the BLR. This is in a good agreement if one perceives BLR as a clumpy medium with small filling factor. For the flattened BLR \citep{2009NewAR..53..140G}:

$$\ind{v}{Kepler} > \ind{v}{turb} \gtrsim \ind{v}{inflow},$$

\noindent where $\ind{v}{Kepler}$ is Keplerian velocity, $\ind{v}{turb}$ is the turbulence velocity and $\ind{v}{inflow}$ is the inflow velocity.

\begin{figure}
   \centering
   \includegraphics[width=\hsize]{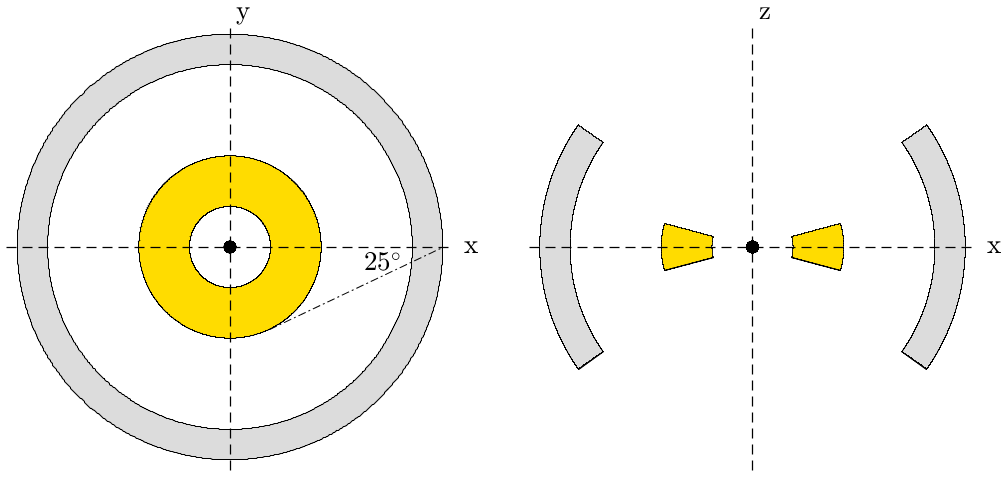}
      \caption{Cartoon showing the model geometry of the BLR (yellow) and the scattering disk (grey) in the face-on (left) and edge-on (right) view.}
         \label{f1}
\end{figure}

In our model, SR is a radially thin region as we assume that the light is being scattered dominantly due to free electrons (Thomson scattering) in the innermost part of the torus. We assume that the electron density is decreasing radially outwards in the form of the power law $\ind{n}{e} \propto r^{-1}$. The SR inner radius $\Rblr{in}{SR}$ is found from the IR reverberation mapping \citep{2011A&A...536A..78K,2014ApJ...788..159K}. The SR outer radius $\Rblr{out}{SR}$ was chosen such that BLR half opening angle when viewed from the edge of the SR is $25^\circ$. Investigation by \citet{2012A&A...548A.121M} for the SR with the flared-disk geometry have shown that optically thin SR $(\tau\le 0.1)$ cannot produce sufficient polarization for Type-1 viewing angles. On the contrary, for optical depths $\tau>3$, multiple scattering can occur, resulting in depolarization. For this reason, we chose to set the total optical depth in radial direction to be $\tau=1$. An illustration of the model is shown in Fig.\@ \ref{f1}.

\subsection{Generic models}
We generated four probe models for which the central SMBH has mass of \SI{d6}{}, \SI{d7}{}, \SI{d8}{} and \SI{d9}{\solarmass}. We expect that the BLR distance from the center increases when the mass of the central SMBH increases, since the mass of the SMBH scales very well with the luminosity of the AGN \citep{2000ApJ...543L.111L,2001MNRAS.327.1111G}. In order to determine the size and the position of the BLR as well as the SR for our probe models, we compiled 14 AGNs for which BH masses and inner radii of the BLR and SR are known from reverberation mapping (see Table \ref{t:objects}).

\begin{table*}
\caption{List of objects with known $\log{\ind{M}{bh}}$, $\ind{L}{5100}$, $\Rblr{in}{BLR}$ and $\Rblr{in}{SR}$ that we used for models. Mass was estimated from reverberation mapping using the $\mathrm{H\beta}$ line and for $<f> = 4.3$ \citep{2013ApJ...773...90G}.}
\centering
\begin{tabular}{lcccccrcccc}
\hline\hline
Object & z & $\log{\ind{M}{bh}}$ & $\log{\ind{L}{5100}}$ & Ref. & $\Rblr{in}{BLR}$ & Ref. & $\Rblr{in}{SR}$ & Ref.\\
	& & \SI{}{\solarmass}	&     	& \SI{}{erg\per\second}	& 	& light days	& 	& light days& \\
\hline
Mrk335		& 0.02579 	& 7.230111	& $43.71 \pm 0.06$ 	&  1	& $14.10 \pm 1.20$ 	&  3, 4	& 141.7 & 7\\
Mrk590		& 0.02639 	& 7.569731	& $43.42 \pm 0.07$ 	&  1	& $25.50 \pm 6.00$  	&  3	& 34.7 	& 7\\
Ark120		& 0.03271 	& 8.068022	& $43.78 \pm 0.07$ 	&  1	& $32.70 \pm 3.00$	&  3	& 428.8 & 7, 8\\
Mrk79		& 0.02219 	& 7.611875	& $43.61 \pm 0.04$ 	&  1	& $29.30 \pm 14.30$ 	&  3	& 69.5 	& 7\\
PG0844+349	& 0.06400 	& 7.858308	& $44.24 \pm 0.04$ 	&  6	& $12.20 \pm 5.20$	&  3	& 104.3 &7 \\
Mrk110		& 0.03529 	& 7.292445	& $43.60 \pm 0.04$ 	&  1	& $26.90 \pm 7.00$	&  3	& 104.3 &7 \\
NGC3227		& 0.00386 	& 6.774994	& $42.24 \pm 0.11$ 	&  1	& $4.40 \pm 0.40 $	&  3	& 25.0 	& 7, 8\\
NGC3516		& 0.00884 	& 7.394509	& $42.73 \pm 0.21$ 	&  1	& $14.60 \pm 1.30$	&  3	& 61.2 	& 7\\
NGC4051		& 0.00234 	& 6.129727	& $41.96 \pm 0.20$ 	&  1	& $2.50 \pm 0.20 $	&  3	& 38.0 	& 7, 8\\
NGC4151		& 0.00332 	& 7.555236	& $42.09 \pm 0.22$ 	&  1	& $6.00 \pm 0.40 $	&  3	& 44.0 	& 7, 8\\
3C273		& 0.15834	& 8.838866	& $45.90 \pm 0.02$ 	&  1	& $306.8 \pm 90.9 $	&  3	& 963 	& 8\\
NGC4593		& 0.00900 	& 6.882240	& $42.87 \pm 0.18$ 	&  1	& $4.50 \pm  0.65$	&  3	& 43.0 	& 7\\
NGC5548		& 0.01718 	& 7.718341	& $43.23 \pm 0.10$ 	&  1, 2	& $17.60 \pm 8.86$	&  3, 5	& 60.0 	& 7\\
Mrk817		& 0.03146 	& 7.586162	& $43.68 \pm 0.05$ 	&  1	& $21.20 \pm 14.70$	&  3	& 180.0 & 7\\
PG1613+658	& 0.12900 	& 8.338928	& $44.71 \pm 0.03$ 	&  1	& $35.00 \pm 15.10  $	&  3, 6	& 595.0 & 7\\
PG1700+518	& 0.29200 	& 8.785679	& $45.53 \pm 0.01$ 	&  1	& $251.80 \pm 42.35$ 	&  3, 6	& 687.0 & 7\\
\hline
\end{tabular}
\tablebib{Optical luminosities are taken from (1) \cite{2013ApJ...767..149B}, (2) \cite{2013ApJ...779..109P}. The estimates for $\Rblr{in}{BLR}$ are taken from (3) \cite{2011ApJ...735...80Z}, (4) \cite{2012ApJ...755...60G}, (5) \cite{2013ApJ...773...90G}, (6) \cite{2000ApJ...533..631K}, The estimates for $\Rblr{in}{SR}$ are taken from (7) \cite{2014ApJ...788..159K}, (8) \cite{2011A&A...536A..78K}.}
\label{t:objects}
\end{table*}

\begin{figure*}
   \centering
   \includegraphics[width=\hsize]{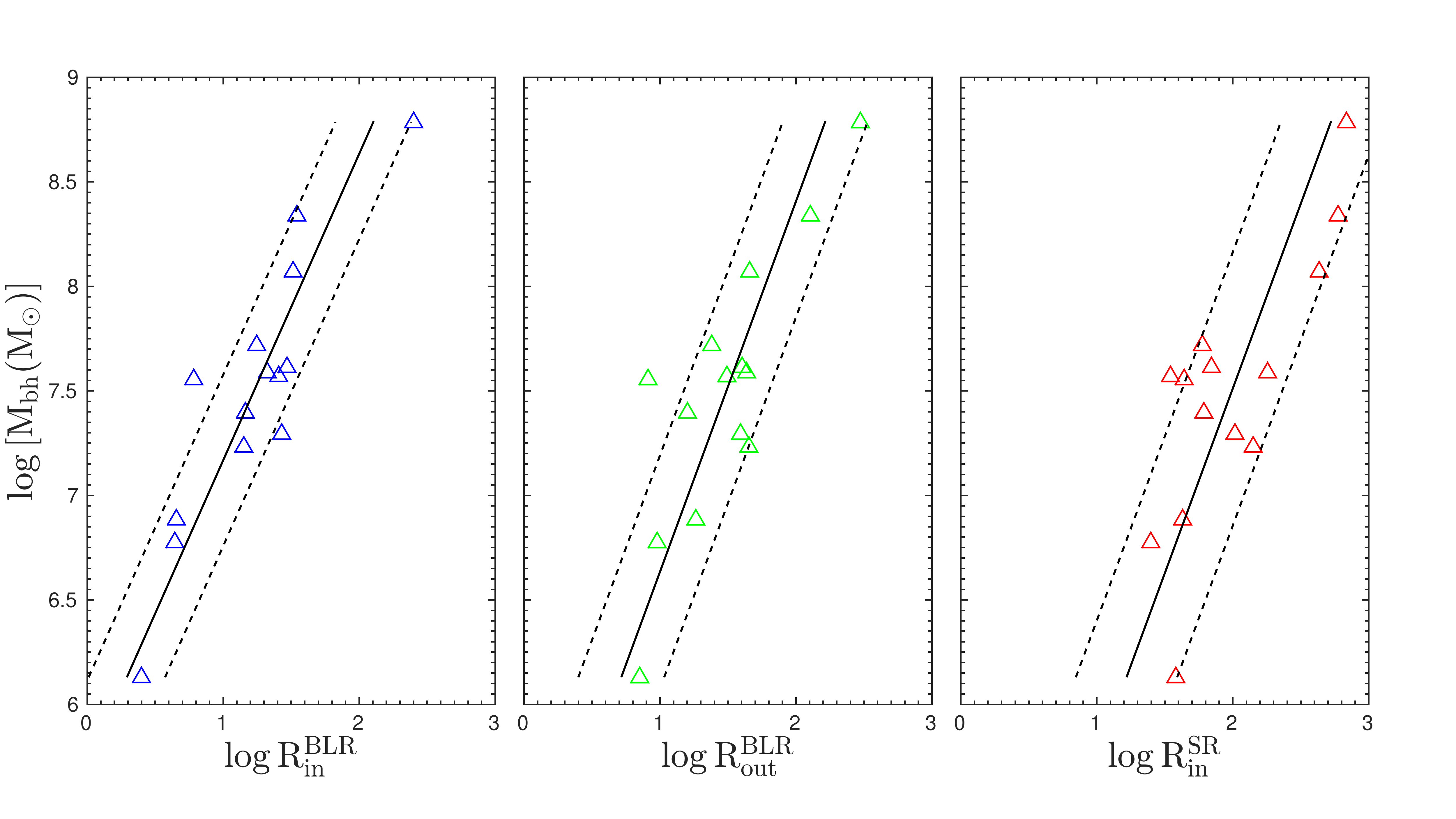}
      \caption{Mass -- radius relation, for $\Rblr{in}{BLR}$ (left panel), $\Rblr{out}{BLR}$ (middle panel) and $\Rblr{in}{SR}$ (right panel). Data taken from literature (see Table \ref{t:objects}) are denoted as triangles, while solid lines represent the best fit. Dashed lines mark the $1\sigma$ uncertainty.}
   \label{f2}
\end{figure*}

We fitted $\ind{M}{bh}$ -- radius relation with a power law in the form:

\begin{equation}
\label{eq:fit}
 \log{\ind{M}{BH}} = C_1\log{R} + C_2,
\end{equation}
\noindent where $R$ takes the values for $\Rblr{in}{BLR}$, $\Rblr{out}{BLR}$ and $\Rblr{in}{SR}$. In Fig.\@ \ref{f2}, we show mass-radius relationship with 1$\sigma$ uncertainty. Fit constants are listed in the Table \ref{t:fit}. This way we obtained a rough estimate of the BLR and SR sizes for our model setup. We represent the goodness of fit using the adjusted coefficient of determination $\bar{R}^2$.

\begin{table}
\caption{The constants $C_1$ and $C_2$ for the mass--radius relation (eq.\, \ref{eq:fit}, second and third column). Adjusted coefficient of determination $\bar{R}^2$ (the last column) for the performed fit.}
\centering
\begin{tabular}{cccc}
\hline\hline
$R$ & $C_1$ & $C_2$ & $\bar{R}^2$ \\
\hline
$\Rblr{in}{BLR}$  & $0.682 \pm 0.096$ &  $-3.890 \pm 0.723$ & 0.7915\\
$\Rblr{out}{BLR}$ & $0.564 \pm 0.108$ &  $-2.743 \pm 0.812$ & 0.6690\\
$\Rblr{in}{SR}$   & $0.566 \pm 0.127$ &  $-2.248 \pm 0.958$ & 0.5899\\
\hline
\end{tabular}
\label{t:fit}
\end{table}

With known fit constants, we generated values for the $\Rblr{in}{BLR}$, $\Rblr{out}{BLR}$, $\Rblr{in}{SR}$ and $\Rblr{out}{SR}$ for the set of four different SMBHs (see Table \ref{t:MR}). Our approach is the following. For each model with known input mass of the SMBH, we solve 3D radiative transfer using \textsc{stokes}. We apply the AP15 method to the simulated results, and finally, we compare the value of the obtained SMBH mass with the value of input SMBH mass.

\begin{table}
\caption{Central SMBH masses, inner and outer radius of the BLR and inner radius of the SR that we used in our model.}
\centering
\begin{tabular}{ccccc}
\hline\hline

\multirow{2}{*}{Mass}	& \multirow{2}{*}{$\Rblr{in}{BLR}$} & \multirow{2}{*}{$\Rblr{out}{BLR}$} & \multirow{2}{*}{$\Rblr{in}{SR}$} & \multirow{2}{*}{$\Rblr{out}{SR}$}\\
& & & &\\
\SI{}{\solarmass} & ld & ld & ld & ld\\
\hline
\SI{d6}{} & 1.597   & 4.385   & 13.968  & 20.262\\
\SI{d7}{} & 7.681   & 16.076  & 51.372  & 74.277\\
\SI{d8}{} & 36.944  & 58.934  & 188.939 & 272.288\\
\SI{d9}{} & 177.700 & 216.043 & 694.893 & 998.170\\
\hline
\end{tabular}
\label{t:MR}
\end{table}

\section{Observations} \label{S.observations}
We have selected four AGNs with prominent change of $\varphi$ across the line profile: NGC4051, NGC4151, 3C273 and PG0844+349. Spectropolarimetry was done with \SI{6}{\meter} telescope of SAO RAS (The Special Astrophysical Observatory of the Russian Academy of Sciences) using modified version of the SCORPIO spectrograph (see \citealt{2005AstL...31..194A,2011BaltA..20..363A}). Data reduction and the method of calculating the polarization parameters as well as corrections for the interstellar polarization is done in the same way as it is described in \cite{2012AstBu..67..438A}. Model parameters for these objects are given Table \ref{t:obse}.
In order to test the AP15 method theoretically, we modeled each of these objects using observational data available from the literature. This is important since we can perform direct comparison of the results obtained from the model with the newest spectropolarimetric observations using the SAO RAS \SI{6}{\meter} telescope. $\Rblr{in}{BLR}$ and $\Rblr{in}{SR}$ were taken from the literature using dust reverberation method \citep[][Table \ref{t:objects}]{2011A&A...536A..78K,2014ApJ...788..159K}, while $\Rblr{out}{BLR}$ and $\Rblr{out}{SR}$ were computed in the same way as for the generic models. Model parameters for these objects are given in Table \ref{t:obse}. Input mass was obtained by applying the AP15 method to the observational data.

\textbf{NGC4051} is a relatively nearby Seyfert 1 galaxy with the cosmological redshift equals to 0.0023, known for its highly variable X-ray flux \citep{2004MNRAS.348..783M}. NGC 4051 was extensively observed in the high energy band to see if the rapid continuum variations observed in the X-ray spectra are correlated to the optical band fluctuations. This is not the case, despite that the time-averaged X-ray and optical continuum fluxes are well correlated. Only the flux of the broad $\mathrm{H\beta}$ line is lagging behind the optical continuum variations by 6 days, allowing us to estimate the mass of the central supermassive black hole \citep{2000ApJ...542..161P}. The optical continuum polarization of NGC 4051 was measured by \citet{1983ApJ...266..470M} and \citet{2002MNRAS.335..773S}, who found a polarization degree of $0.52 \pm 0.09\%$ and $0.55 \pm 0.04\%$, respectively. The polarization position angle was found to be parallel to the radio axis of the AGN, such as expected for most of Type-1 objects \citep{1993ARA&A..31..473A}.

\textbf{NGC4151} is a 1.5 Seyfert galaxy situated at $z=0.0033$ \citep{1991S&T....82Q.621D}, which is sometimes considered to be the archetypal Seyfert 1 galaxy \citep{2008A&A...486...99S,2010A&A...509A.106S}. It is one of the brightest Type-1 AGN in the X-ray and ultraviolet band, and its bolometric luminosity is of the order of \SI{5d43}{erg\per\second} \citep{2002ApJ...579..530W}. The mass of its central supermassive black hole was estimated by optical and ultraviolet reverberation techniques and is estimated at $~\SI{4.5d7}{\solarmass}$ \citep{2006ApJ...651..775B}. Since NGC 4151 stands out thanks to its high fluxes and proximity, its optical polarization was extensively observed (see \citealt{2016A&A...591A..23M}). The averaged \SI{4000}{}--\SI{8000}{\angstrom} continuum polarization is below 1\%, with a polarization position angle parallel to the radio axis of the system. In the optical range, NGC 4151 shows flux variations of the continuum and of the broad lines up to a factor of ten or greater \citep{2008A&A...486...99S,2010A&A...509A.106S}. The wings of broad lines also vary greatly from very intensive ones corresponding to type Sy1 in the maximum state of activity to almost complete absence in the minimum state of activity. In April 1984, the nucleus of NGC 4151 went through a very deep minimum and the broad wings of hydrogen lines almost completely vanished and the spectrum of the nucleus was identified as a Sy 2 \citep{1984MNRAS.211P..33P}. In this phase, the intensity in the broad component of a spectral line is too weak that probably the AP15 method could not be used. Another technique must be used to explore the geometry of the object, such as proved by \citet{2017A&A...604L...3H} and \citet{2017A&A...607A..40M}.

\textbf{3C273} is a well known flat-spectrum radio source quasar with broad emission lines. It is the brightest and one of nearest quasars known to us ($z=0.158$, \citealt{1987A&A...176..197C,1990A&A...234...73C}). It is a radio-loud object, i.e. its radio-to-millimeter energy output is dominated by synchrotron emission from a kilo-parsec, one-sided jet, whose emission extends up to the infrared and optical bands. 3C 273 is particularly bright in the optical and ultraviolet domains, which enabled the detection of the polarization of the optical emission. Its mean optical core polarization was measured by \citet{1968ApJ...151..769A} and is of the order of $0.2 \pm 0.2\%$, being consistent with galactic interstellar polarization \citep{1966ZA.....64..181W}. The optical polarization emerging from the jet structure is higher as resolved into a number of highly polarized knot structures by \citet{1993Natur.365..133T}. Nevertheless, Balmer emission lines were first measured by \citet{1963Natur.197.1040S} and allowed a determination of the central black hole mass from the average line profiles \citep{2000ApJ...533..631K}. 

\textbf{PG0844+349} is a radio-quiet quasar at a cosmological redshift of $z=0.064$ that, in comparison to most quasars, was not originally detected in the radio frequency: at radio wavelengths, its nucleus is unresolved \citep{1994AJ....108.1163K}. It was first discovered in the Palomar Green sample \citep{1983ApJ...269..352S} and found to possess strong Fe II emission and weak forbidden narrow lines, a behavior that is expected from narrow-line Seyfert-1s (NLS1). On the other hand, the X-ray properties of PG 0844+349 are aligned with the NLS1 classification (a steep soft X-ray spectrum and strong variability, see \citealt{2001ASPC..251..112B}), and the optical polarization measurements achieved by \citet{2011AstL...37..302A} also point towards a regular NLS1 object (optical continuum polarization of $0.85 \pm 0.10 \%$). Hence, using Type-1 AGN reverberation mapping techniques, \citet{2004ApJ...613..682P} estimated the mass of the central black hole to be $(9.24 \pm 3.81)\times\SI{d8}{\solarmass}$. 

\begin{table*}
\caption{\label{t2} Central SMBH masses, inner and outer radius of the BLR and SR that we used in our model for comparison with the observed data.}
\centering
\begin{tabular}{lccccc}
\hline\hline
\multirow{2}{*}{Object}		& \multirow{2}{*}{$\log(\ind{M}{POL}/\ind{M}{\odot})$} & \multirow{2}{*}{$\Rblr{in}{BLR}$} & \multirow{2}{*}{$\Rblr{out}{BLR}$} & \multirow{2}{*}{$\Rblr{in}{SR}$} & \multirow{2}{*}{$\Rblr{out}{SR}$} \\
& & & & &\\
& & ld & ld & ld & ld\\
\hline
NGC 4051		& $6.69 \pm 0.21$ & 4.3 	& 15.0 	& 38.1 	& 53.7\\
NGC 4151		& $7.21 \pm 0.27$ & 6.6 	& 17.5 	& 44.0 	& 63.8\\
3C 273			& $8.85 \pm 0.27$ & 306.8 	& 440.6	& 963.7	& 2035.8\\
PG0844+349		& $7.70 \pm 0.14$ & 12.2 	& 77.4	& 189.0	& 357.6\\
\hline
\end{tabular}
\label{t:obse}
\end{table*}


\section{Results} \label{S.results}
We present our results which can be divided into two parts, first, the results of modeling, and second we compare our models with observations. The same convention used by \citet{2007A&A...465..129G} was adopted in this work. Namely, $\varphi$ is parallel to the symmetry axis of the model when $\varphi=\ang{90}$, which was observed for Type-1 objects, or $\varphi$ is orthogonal to the symmetry axis when $\varphi=\ang{0}$, that is again, observed for Type-2 objects.
\subsection{Generic modeling}
We simulated different geometries of the BLR. First we performed the simulation for different masses of the black holes with assumption of a pure Keplerian motion, after that we consider the radial inflow and vertical outflow as additional components in gas motion to the Keplerian caused by the black hole mass. We simulated both cases where Keplerian motion is in anticlockwise (positive) and clockwise (negative) direction.

\subsubsection{Pure Keplerian gas motion in the BLR}
\begin{figure*}
    \centering
    \includegraphics[width=\hsize]{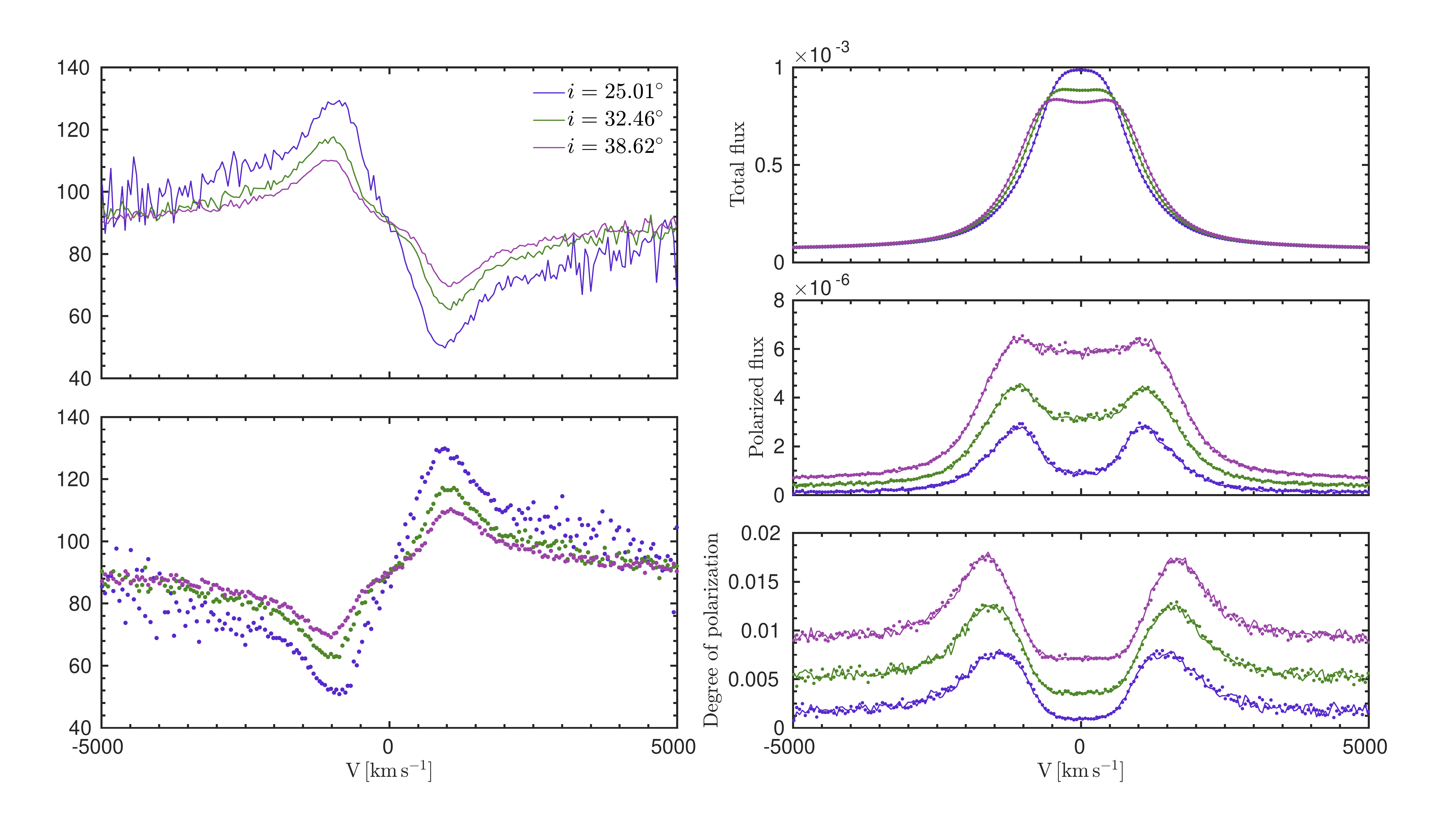}
    \caption{On the left panels, modeled polarization plane position angle $\varphi$ is shown when the system is rotating anticlockwise (top) or when rotating in clockwise direction (bottom), total unpolarized flux (TF, top right), polarized flux (PF, middle right), degree of polarization (PO, bottom right). SMBH has mass of \SI{d6}{\solarmass}. We plot the results in solid lines for three viewing inclinations: $i = \ang{25.01},\ \ang{32.46},\ \mathrm{and}\ \ang{38.62}$ respectively, while dotted lines represent the results for the opposite direction of rotation. Note the symmetry of $\varphi$ with respect to the continuum level due to the opposite direction of rotation. Opposite direction of rotation does not affect TF, PF and PO. Total and polarized fluxes are given in arbitrary units.}
    \label{e6_2_TF}%
  \end{figure*}

\begin{figure*}
    \centering
    \includegraphics[width=\hsize]{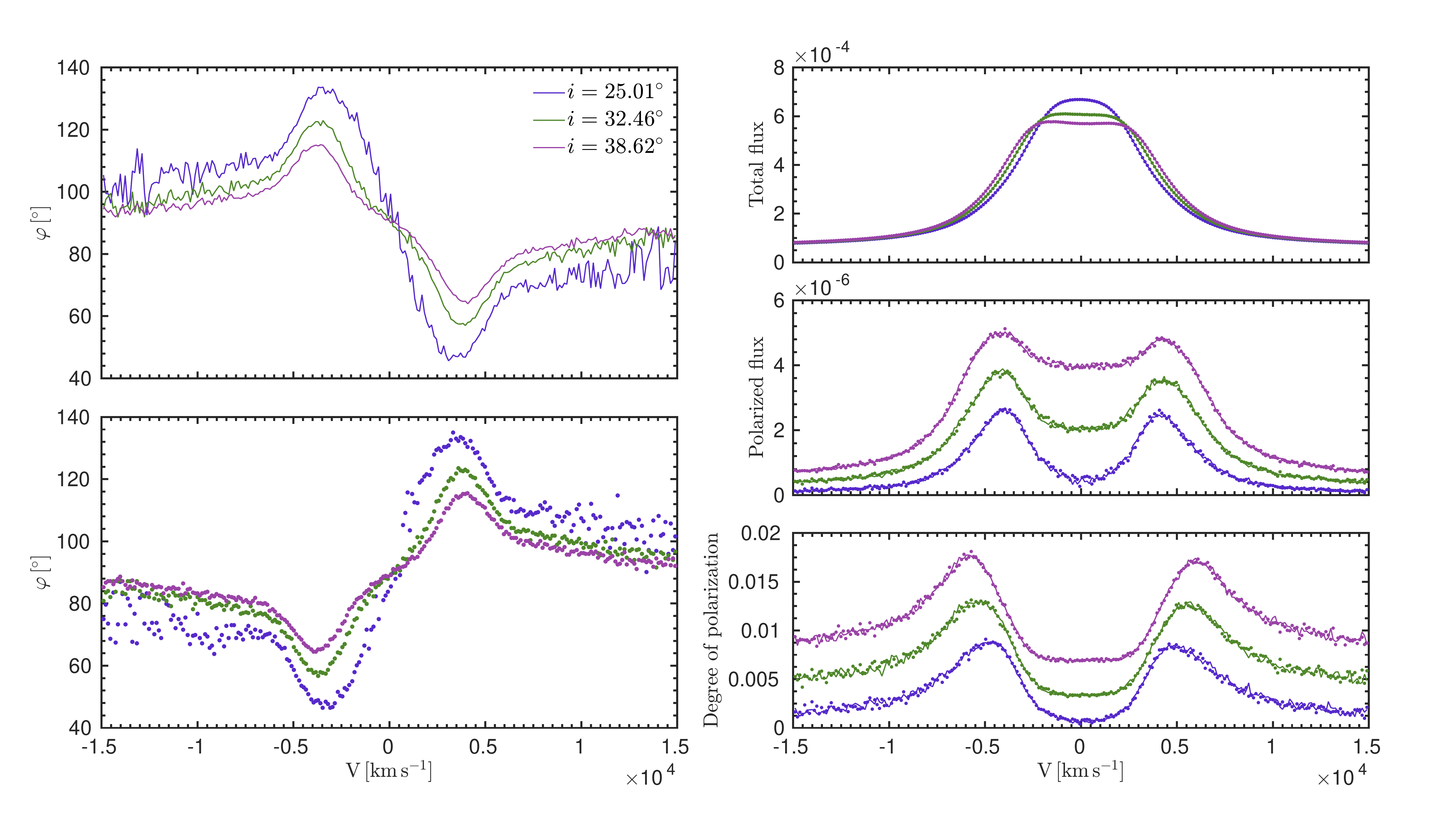}
    \caption{Same as Fig.\@ \ref{e6_2_TF}, but for SMBH of \SI{d9}{\solarmass}}
    \label{e9_18_TF}%
  \end{figure*}

We consider the pure Keplerian motion of the BLR emitting gas, taking that there are no other effects as e.g. outflows and inflows. We present the results of the four probe models. In Figs.\@ \ref{e6_2_TF} and \ref{e9_18_TF}, we show the simulated profiles of $\varphi$, polarized flux (PF), degree of polarization (PO) and total flux (TF) across the broad line profile.

\noindent Each scattering element can see the velocity resolved BLR emission which produces polarized lines that are broader than the unpolarized lines. Simulated degree of polarization is of the same order of magnitude as the one obtained from observations and is typically around 1\% or less \citep{2016A&A...591A..23M}. From our models (Figs \ref{e6_2_TF} and \ref{e9_18_TF}, bottom right panels), we can see that the degree of polarization is sensitive to inclination. Extensive modeling with complex radiation reprocessing \citep[see e.\,g.][for more details]{2012A&A...548A.121M} have shown that the total PO is increasing as we start looking from the face-on viewing angle towards Type-2 viewing angles. Although we included only equatorial scattering in our model, the dependence of PO with inclination is following this trend. The PO profile peaks in the line wings and reaches minimum in the line core just as it was shown by \citet{2005MNRAS.359..846S}. This feature was very well observed for the case of Mrk 6 \citep{2002MNRAS.335..773S, 2014MNRAS.440..519A} and it is supporting the suggested scattering geometry.

The polarization plane position angle is aligned with the disk rotation axis, hence also with the radio jet axis. In Figs.\@ \ref{e6_2_TF}, \ref{e9_18_TF} (left panels), we show the simulated profiles of $\varphi$ for three viewing inclinations. The $\varphi$ profiles show typical symmetric swing that was predicted for Type-1 objects where the radiation from the Keplerian rotating disc-like BLR is being scattered by outer dusty torus \citep[][AP15]{2005MNRAS.359..846S}. The direction of rotation only affect $\varphi$, while TF, PF and PO remains unaffected. For anticlockwise rotation, $\varphi$ reaches maximum value in the blue part of the line and minimum in the red part of the line. The $\varphi$ swing occurs around the level of continuum $\ind{\varphi}{c} = \ang{90}$. Due to the symmetry of the model (also for all other models performed in the paper), $\varphi$ is symmetric with respect to the continuum polarization in such a way that for a given inclination $i$, it satisfies the following:

\begin{equation} \label{eq:PA}
 \varphi(\ang{180} - i = \ang{180} - \varphi(i).
\end{equation}
In other words, for a given half-opening angle of the torus $\theta_0$, and for Type-1 inclinations where $0\le i\le\ang{90}-\theta_0$, observer can see one way of rotation, and the corresponding $\varphi$ profile will be as shown in Figs. \ref{e6_2_TF} and \ref{e9_18_TF}. If the system is viewed for Type-1 viewing angle where $\ang{90}+\theta_0 \le i\le \ang{180}$, opposite direction of rotation is observed and the resulting $\varphi$ satisfies Eq. \ref{eq:PA}. This symmetry can be seen in Figs. \ref{e6_2_TF} and \ref{e9_18_TF} (left panels). Thus spectropolarimetric observations of Type-1 Seyferts can disentangle the rotation direction of the gas by observing the $\varphi$ profile.
The Eq. \ref{eq:PA} is satisfied within the Monte Carlo uncertainty and it was used for improving photon statistics in our simulations by a factor of 2, by simply taking average value of the STOKES parameters for inclinations $i$ and $\ang{180} - i$.

\begin{figure*}
    \centering
    \includegraphics[width=\hsize]{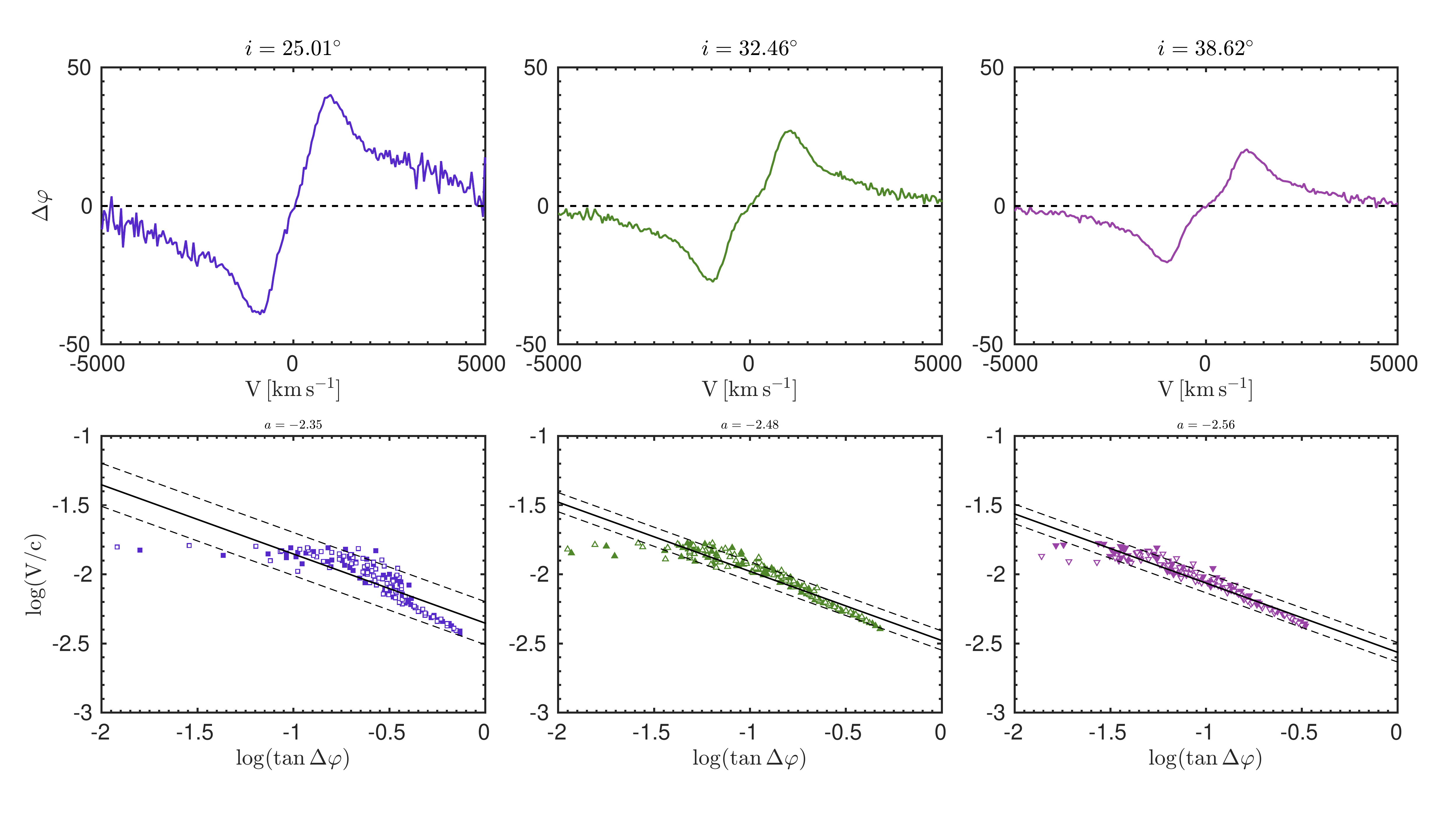}
    \caption{Modeled polarization $\varphi$ (upper panels) and velocities (lower panels) across $\mathrm{H}\alpha$ profiles for the model with central mass of \SI{d6}{\solarmass}. Filled symbols are for the blue part of the line and open symbols are for the red part of the line. Solid line represents the best fit.}
    \label{e6_masfit}%
   \end{figure*}

When performing AP15 method to the modeled data, one needs to consider polarization only in the broad line and for that, it was necessary to subtract the continuum polarization for all Type-1 inclinations:
\begin{equation}
 \Delta\varphi = \varphi - \ang{90}.
\end{equation}
Since all of our observed objects are rotating clockwise (see Sect. \ref{S.comparison}), we performed the AP15 method assuming opposite direction of rotation, without introducing new simulations. In Fig.\@ \ref{e6_masfit} (lower panels), we show the fit described by AP15 method. We find that Keplerian motion can be traced across the $\varphi$ profile for Type-1 viewing inclinations. The region inside the $1\sigma$ error around the linear fit is becoming smaller as we go from face-on towards edge-on inclinations. For inclinations $25^\circ$ or lower, the simulated data show much higher scatter around the straight line rather than for the cases with an average inclination.

\begin{table}
\caption{Input mass (Column 1), viewing inclinations (Column 2) and masses obtained for probe models (Column 3). Masses are given in \SI{}{\solarmass}.}
\centering
\begin{tabular}{ccc}
\hline\hline
$\log{\ind{M}{input}}$ & $i(^\circ)$ & $\log(\ind{M}{MOD}/\ind{M}{\odot})$ \\
\hline
             & 25.01 & $6.72 \pm 0.10$\\
6            & 32.46 & $6.44 \pm 0.06$\\
             & 38.62 & $6.28 \pm 0.05$\\
\hline
             & 25.01 & $7.59 \pm 0.10$\\
7            & 32.46 & $7.30 \pm 0.07$\\
             & 38.62 & $7.17 \pm 0.04$\\
\hline
             & 25.01 & $8.65 \pm 0.11$\\
8            & 32.46 & $8.39 \pm 0.07$\\
             & 38.62 & $8.23 \pm 0.06$\\
\hline
             & 25.01 & $9.67 \pm 0.16$\\
9            & 32.46 & $9.43 \pm 0.12$\\
             & 38.62 & $9.27 \pm 0.10$\\
\hline
\end{tabular}
\label{t3}
\end{table}

   \begin{figure*}
    \centering
    \includegraphics[width=\hsize]{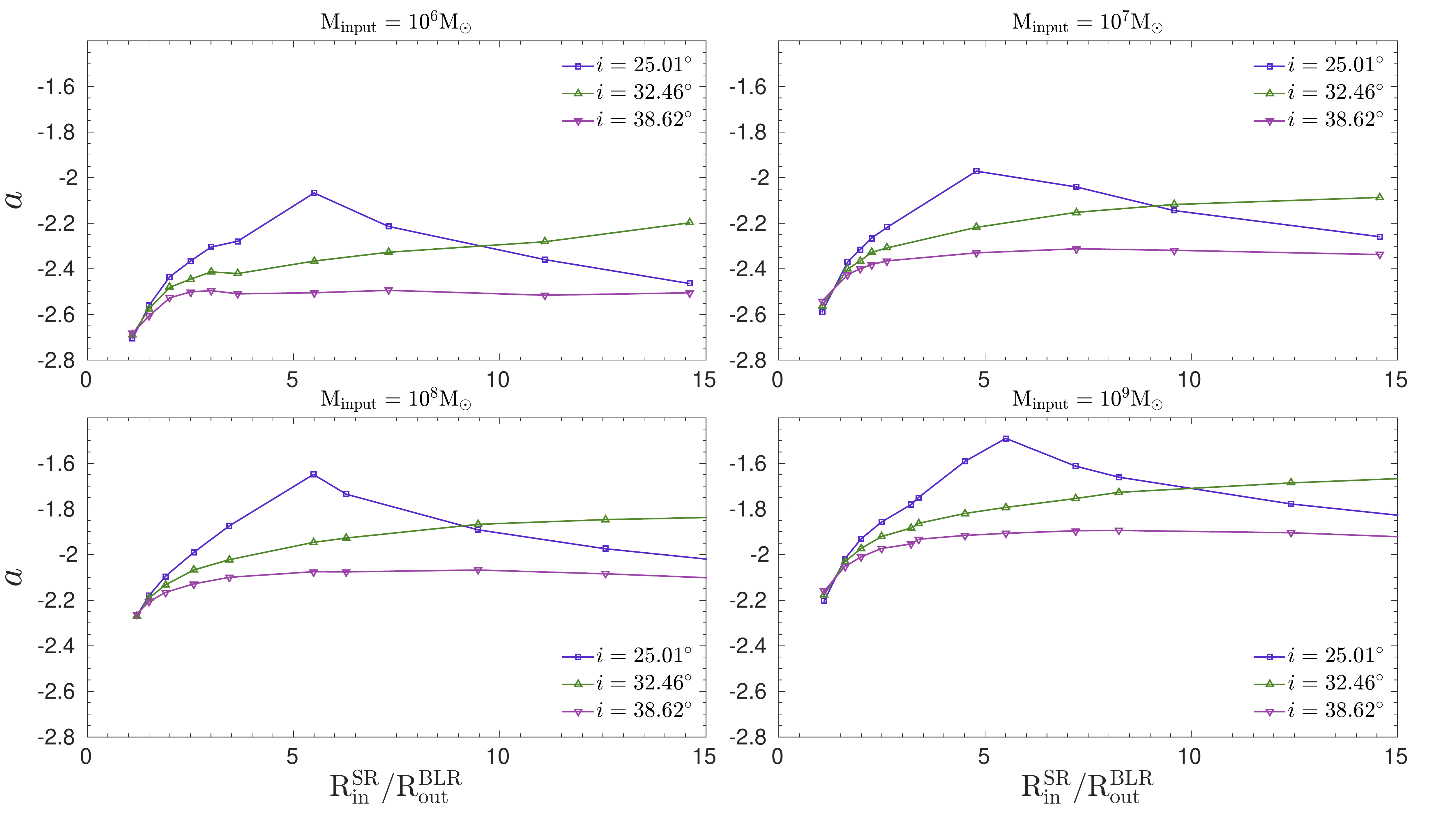}
    \caption{Dependance of the parameter $a$ on the ratio between the inner radius of the SR ($\Rblr{in}{SR}$) and the outer radius of the BLR ($\Rblr{out}{BLR}$) for three given inclinations.}
    \label{a}%
   \end{figure*}

   \begin{figure*}
    \centering
    \includegraphics[width=\hsize]{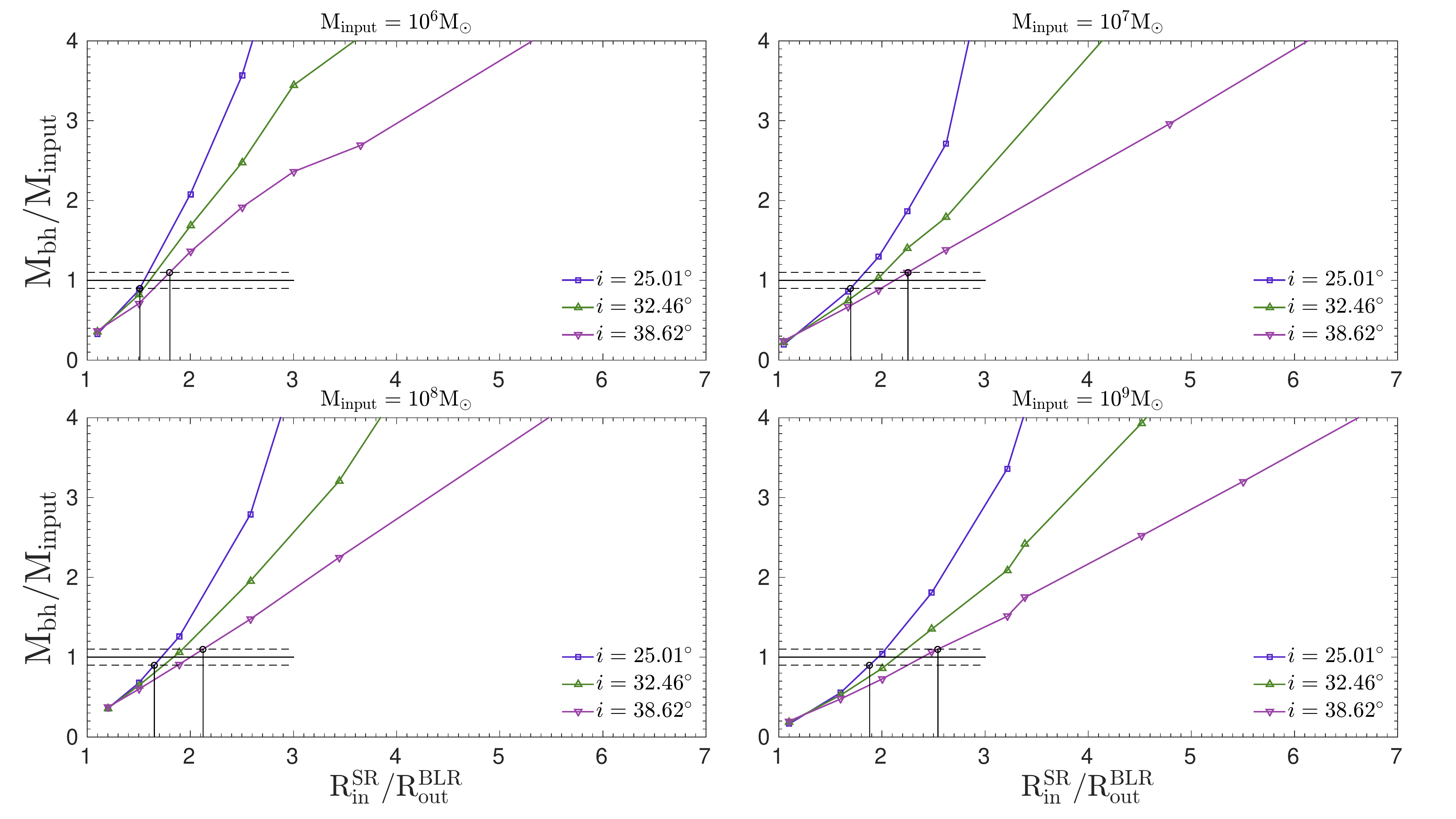}
    \caption{Black hole mass estimation as a function of the ratio between the inner radius of the SR ($\Rblr{in}{SR}$) and the outer radius of the BLR ($\Rblr{out}{BLR}$) for three given inclinations. Horizontal dashed lines represent the interval of 10\% deviation from the input mass (solid line).}
    \label{Mbh}%
   \end{figure*}

The effect of a wide SR (in our case $\theta_0 = 35^\circ$) can lead to a mass estimates by a maximum of $\sim 1.5$ higher than the ones obtained for equatorial scattering, and only if the SR lies much farther away from the BLR (see Eq.\, \ref{eq:Mbh}). In this case, the influence of the viewing inclination must be taken into account.

It is important to note that in the Eq.\,\ref{eq:Mbh}, we used the inner radius of the SR in order to estimate the mass of the SMBH. However, the SR is not acting as a mirror from which the light is being scattered from the inner wall. Scattering events occur in the entire SR, and they all contribute to the total $\varphi$ shape. Obtaining the value for parameter $a$ is a straightforward procedure, but the final estimated SMBH value is largely depending on the actual value of $\ind{R}{sc}$. In the optically thick media, the largest fraction of photons is being scattered of the inner side of the SR.

One of the factors that have significant impact on the $\varphi$ amplitude is the mutual distance between the BLR and the SR \citep{2005MNRAS.359..846S}. The amplitude of $\varphi$ is decreasing when mutual distance increases, which affects black hole mass estimation. Therefore, we investigated different cases with various mutual distances between the BLR and SR while keeping the same thickness and the optical depth of the SR. In Figs.\@ \ref{a} and \ref{Mbh}, we show the influence of different mutual distance between the two regions, and how it affects the parameter $a$ and SMBH estimates. Our models show that mutual distance between the BLR and SR has a great influence on the parameter $a$ which consequently greatly affects our black hole mass estimates. One can see that parameter $a$ shows the same profile and the same inclination dependence for all simulated cases. Only when the SR is adjacent to the BLR we obtain inclination independence of the SMBH mass estimates. Due to the nature of the Eq.\,\ref{eq:Mbh}, SMBH mass estimates are increasing when the mutual distance increase. For a given accuracy of 10\%, we find that the best SMBH estimates for all four cases are when the ratio of the inner radius of SR and the outer radius of the BLR is between 1.5 and 2.5 (Fig.\@ \ref{Mbh}). For the inclinations of $25^\circ$ or less (face-on view), contribution of equatorial scattering is low and we find that Keplerian motion cannot be recovered from the $\varphi$ profile.

\subsubsection{Keplerian motion and radial inflow}

\begin{figure*}
    \centering
    \includegraphics[width=\hsize]{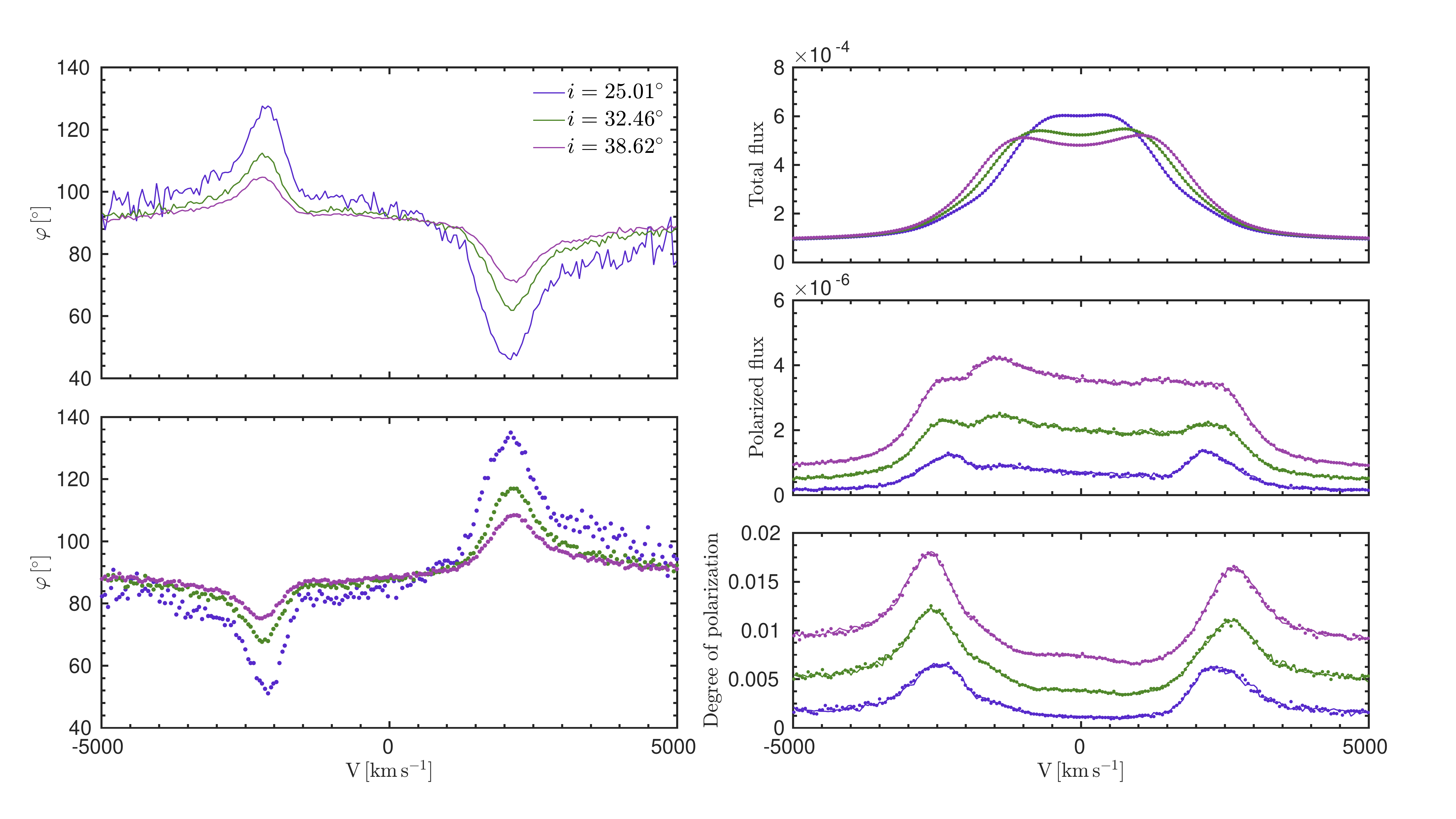}
    \caption{Same as Fig.\@ \ref{e6_2_TF}, but beside Keplerian motion, by large inflow of \SI{2000}{\kilo\meter\per\second} is included in the BLR kinematics.}
    \label{e6_in3_TF}%
  \end{figure*}

We investigated a particular case when the BLR is undergoing a constant radial inflow. We tested three cases with BLR radial inflow velocity equals to \SI{500}{\kilo\meter\per\second}, \SI{1000}{\kilo\meter\per\second} and \SI{2000}{\kilo\meter\per\second}. Similarly as before, in Fig.\@ \ref{e6_in3_TF}, we show simulated profiles of $\varphi$, PF, PO and TF. In this regime, the SR can see additional component of the BLR velocity, which as a net effect increases the absolute value of the radial velocity that a single scattering element can see. This leads to additional line broadening (Fig.\@ \ref{e6_in3_TF}, lower right panel) when compared with the case with pure gas Keplerian motion only. As a resulting effect, the distance between the positions of the maximum and the minimum of the $\varphi$ is increased (Fig.\@ \ref{e6_in3_TF}, left panels). Therefore, for a low velocity radial inflow, mass estimates of the SMBHs are slightly higher than the ones obtained in the case with Keplerian motion only. This overestimation of the SMBH mass mostly affects the model for which the SMBH has mass of \SI{d6}{\solarmass}. For other models, Keplerian motion is even more dominant (except for the very extreme cases which are not expected) and the influence of the radial inflow can be neglected.

\subsubsection{Keplerian motion and vertical outflow}

\begin{figure*}
    \centering
    \includegraphics[width=\hsize]{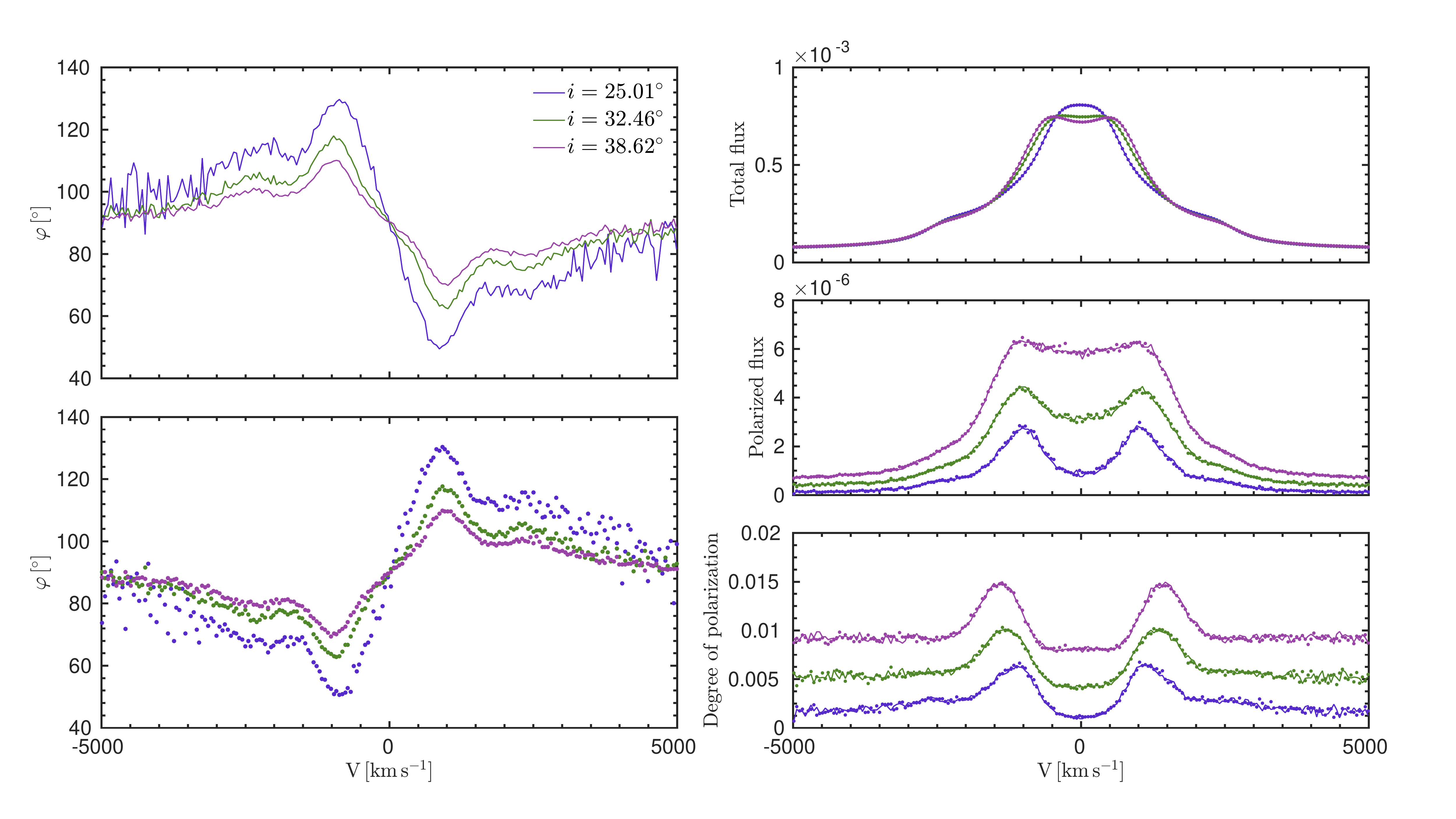}
    \caption{Same as Fig.\@ \ref{e6_2_TF}, except that here the inner one third of the BLR undergoes a constant vertical outflow of \SI{2000}{\kilo\meter\per\second}.}
    \label{e6_out3_TF}%
  \end{figure*}

\begin{figure*}
    \centering
    \includegraphics[width=\hsize]{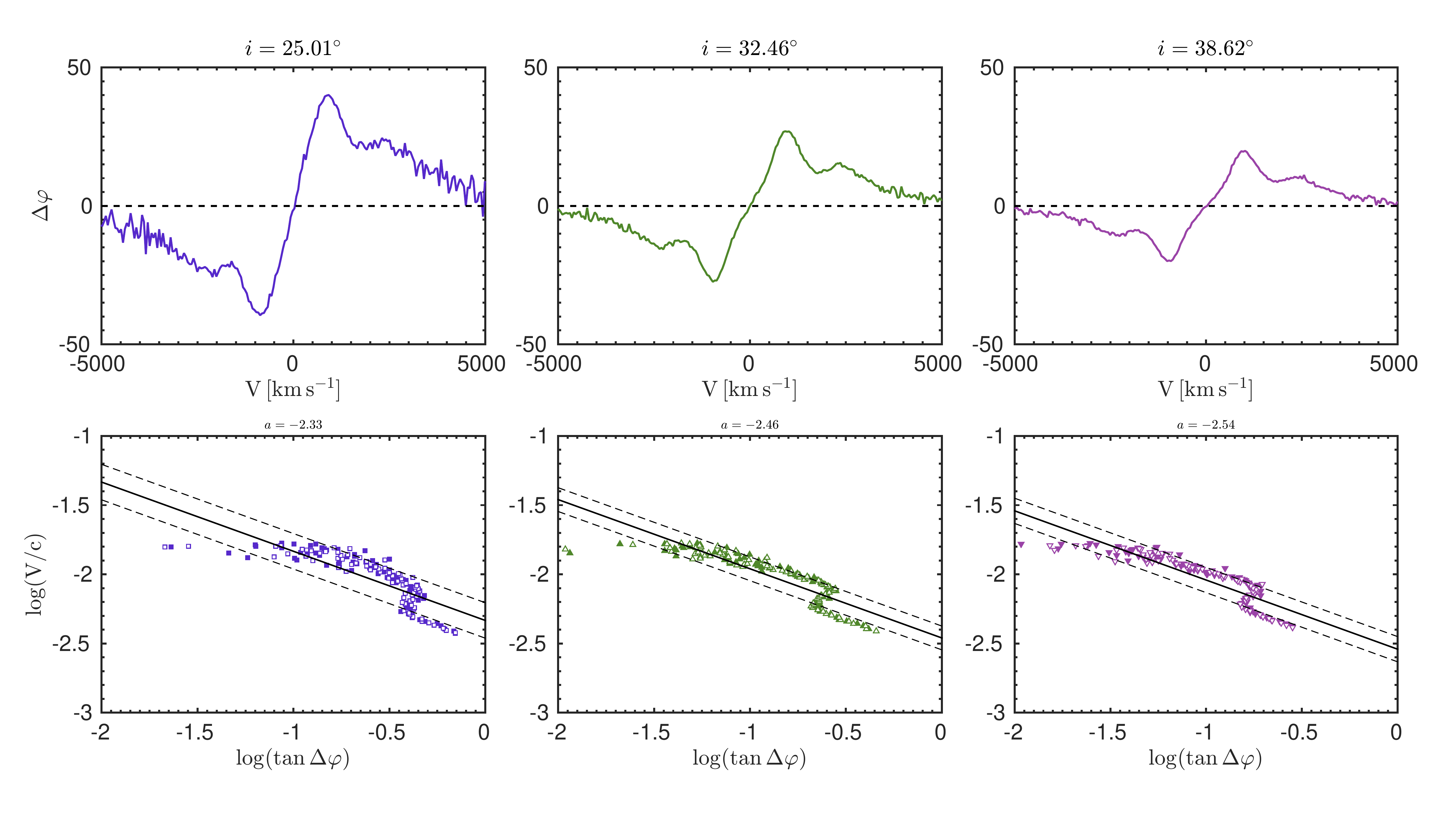}
    \caption{Same as Fig.\@ \ref{e6_masfit}, except that here the inner one third of the BLR undergoes a constant vertical outflow of \SI{2000}{\kilo\meter\per\second}.}
    \label{e6_out3_massfit}%
  \end{figure*}

Another contribution to velocity might be due to vertical outflows. We tested three cases for which the innermost one third of the BLR is undergoing a constant vertical outflow of \SI{500}{\kilo\meter\per\second}, \SI{1000}{\kilo\meter\per\second} and \SI{2000}{\kilo\meter\per\second}. In this case, the equatorial scattering elements will not see this velocity component. Scattering elements above the equatorial plane will see this component multiplied by a factor of $\cos{\alpha}$, where $\alpha$ is latitude, to a maximum of $\cos{35^\circ}$. This can be neglected when the outflow velocity is much lower in comparison with Keplerian velocity. In Fig.\@ \ref{e6_out3_TF}, we show the results of simulated $\bm{\varphi}$, PF, PO and TF influenced by vertical outflows in the BLR of \SI{2000}{\kilo\meter\per\second} for the case where SMBH has the mass of \SI{d6}{\solarmass}. Unpolarized line (bottom right panel) is additionally broadened in the wings. Polarized line (upper right panel) is almost the same as the one for the case with Keplerian motion only (Fig.\@ \ref{e6_2_TF}, upper right panel) due to the reasons explained above. Contribution of outflow velocity is highest for nearly face-on view. In Fig.\@ \ref{e6_out3_massfit}, left panels, the $\varphi$ profile shows additional bump which prevents us from correctly using the AP15 method. We would like to point out again that in our model, the BLR is transparent and that observer can see the radiation coming from both approaching and the receding part of the BLR outflows. We know from observations that this is not the case \citep[e.g\ Mrk 6,][]{2014MNRAS.440..519A}, and we expect to observe radiation from the approaching side of the BLR outflows, while the radiation from the receding side of the BLR outflos should be blocked, thus affecting only the blue part of the line.

\subsection{Comparison with observations} \label{S.comparison}
Here we present the fits of our model with observations in order to estimate the SMBH mass. We fit model data to observational data and compare the results. The results are given in Table \ref{datamodel}, and below we discuss the results and visual comparison for each object.

\begin{figure*}
    \centering
    \includegraphics[width=\hsize]{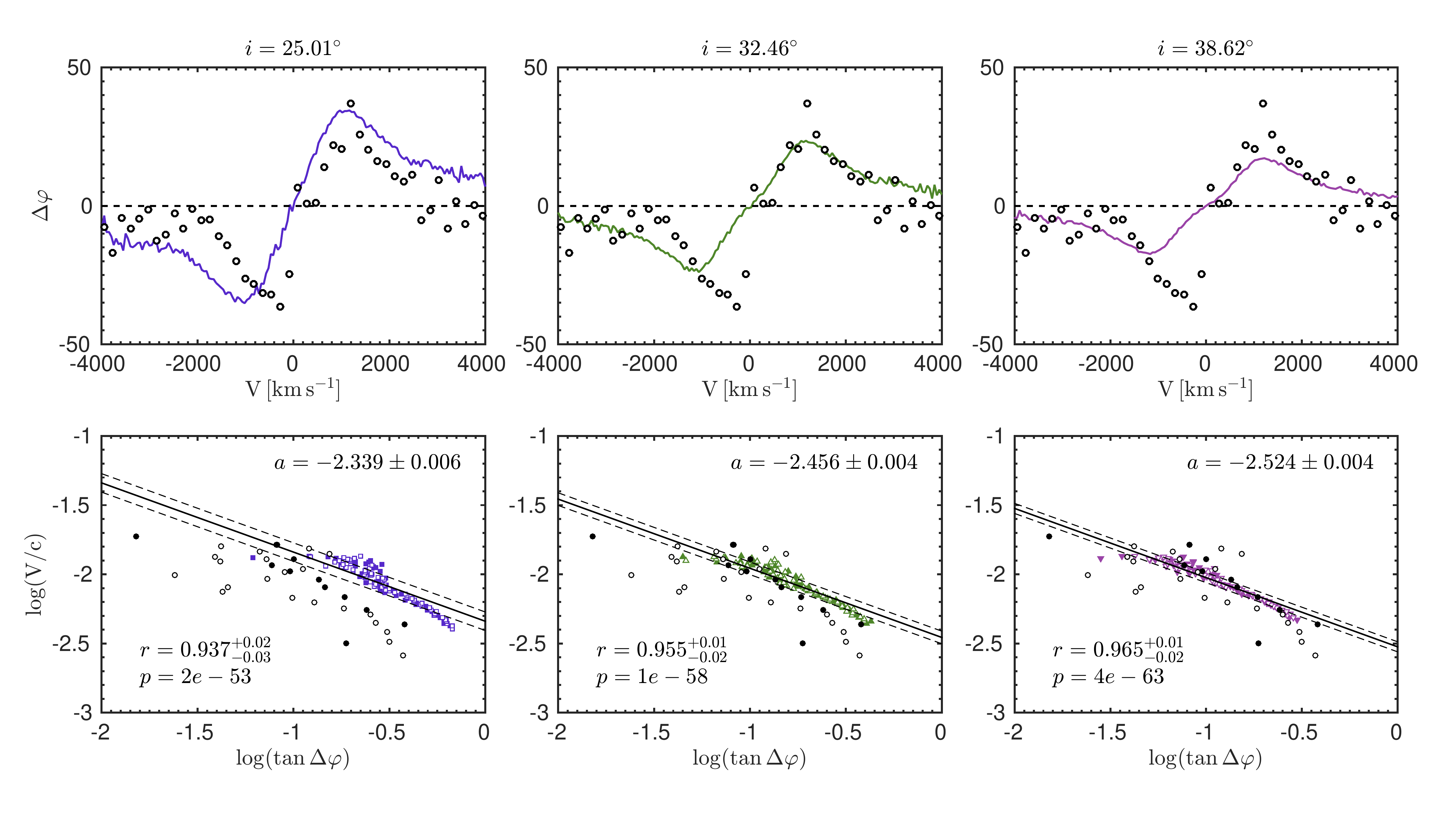}
    \caption{Modeled broad line polarization position angle $\varphi$ (upper panels) and velocities (lower panels) across $\mathrm{H}\alpha$ profiles for the NGC 4051. On the upper panels, data obtained by models is depicted as line while observed data are depicted by empty circles. On the lower panels, filled symbols are for the blue part of the line and open symbols are for the red part of the line for model data. Black circles depict observed data. Solid black line represents the best fit. Values of the parameter $a$, correlation coefficient $r$ and the corresponding $p$ values are shown.}
    \label{NGC4051}%
\end{figure*}

\noindent\textbf{NGC 4051} -- We were able to obtain expected $\varphi$ shape (Fig.\@ \ref{NGC4051}, upper panels).

\noindent The amplitude of the $\varphi$ has very close value as the observed one for the lowest inclination in the model for which $i = \SI{25.01}{^\circ}$. For this inclination, the position of the maximum and the minimum of $\varphi$ is displaced which yields the highest mass estimate. As we start viewing from higher inclinations, the $\varphi$ amplitude is decreasing and we could better fit the line wings (Fig.\@ \ref{NGC4051}, lower panels), and the difference between the estimated values of the SMBH masses and the input mass is smaller. We can see that simulated data in the line wings for $i = 25^\circ$ are deviating from the theoretically predicted straight line. For intermediate inclinations the $1\sigma$ offset is smaller and the fit is better. Similarly, SMBH masses estimated from the fitting of the model data are higher than the input mass as it was in the previous case. For this object, $\Rblr{in}{SR}/\Rblr{out}{BLR}\approx 2.54$.

\begin{figure*}
    \centering
    \includegraphics[width=\hsize]{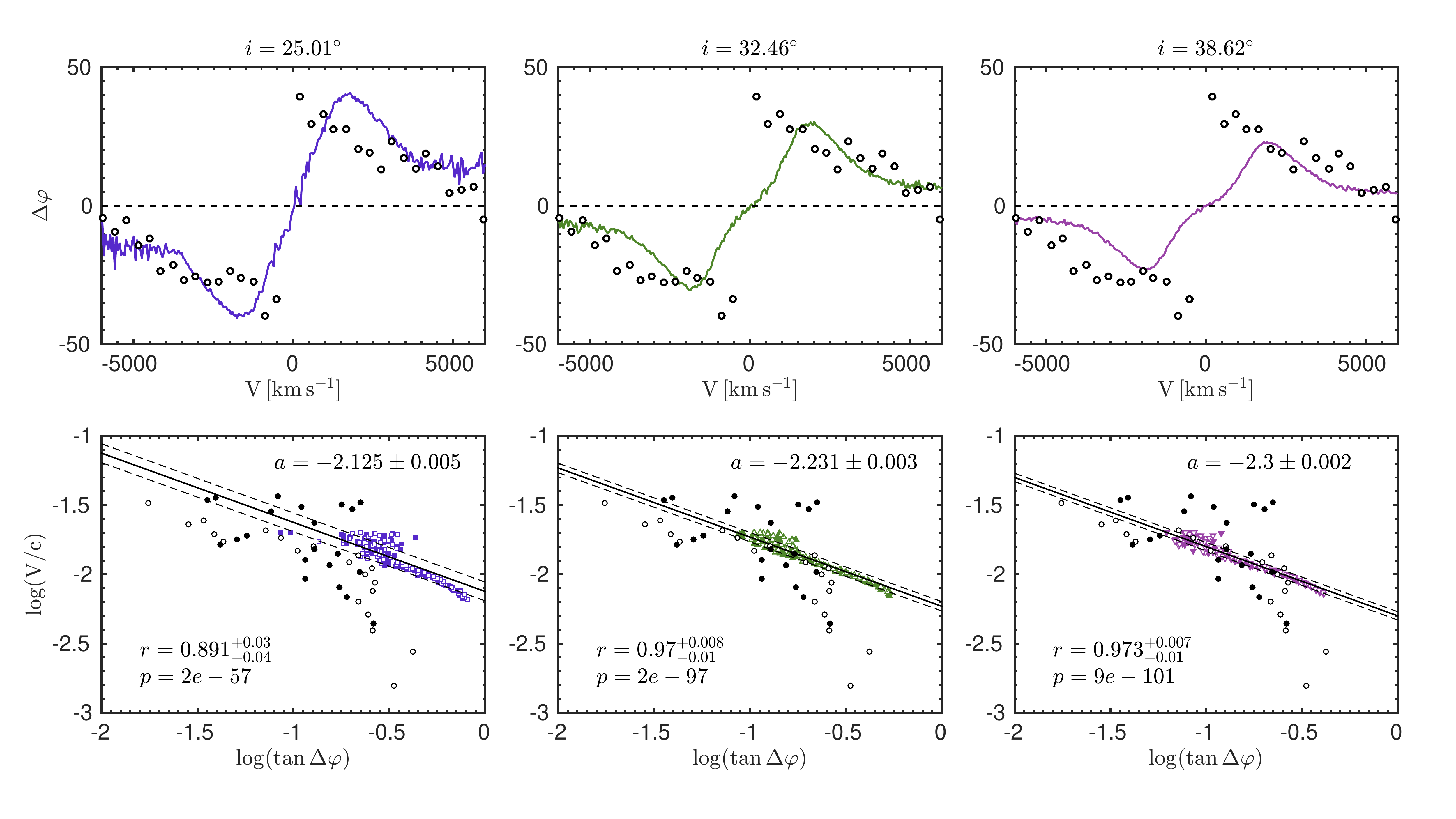}
    \caption{The same as in Fig \ref{NGC4051}, but for NGC4151.}
    \label{NGC4151}%
\end{figure*}

\noindent\textbf{NGC 4151} -- Similar as in the previous case, we obtain the highest mass estimate for the lowest inclination. Keplerian motion is very well shown as a straight line (Fig.\@ \ref{NGC4151}, lower panels), where $1\sigma$ error is small, especially in the case for which $i = 39^\circ$. For this object, $\Rblr{in}{SR}/\Rblr{out}{BLR}\approx 2.51$.

\noindent We can see that in the extreme wings of the line, the modeled $\varphi$ becomes very sensitive to spectral resolution and this sensitivity is smaller for higher inclinations.

\begin{figure*}
    \centering
    \includegraphics[width=\hsize]{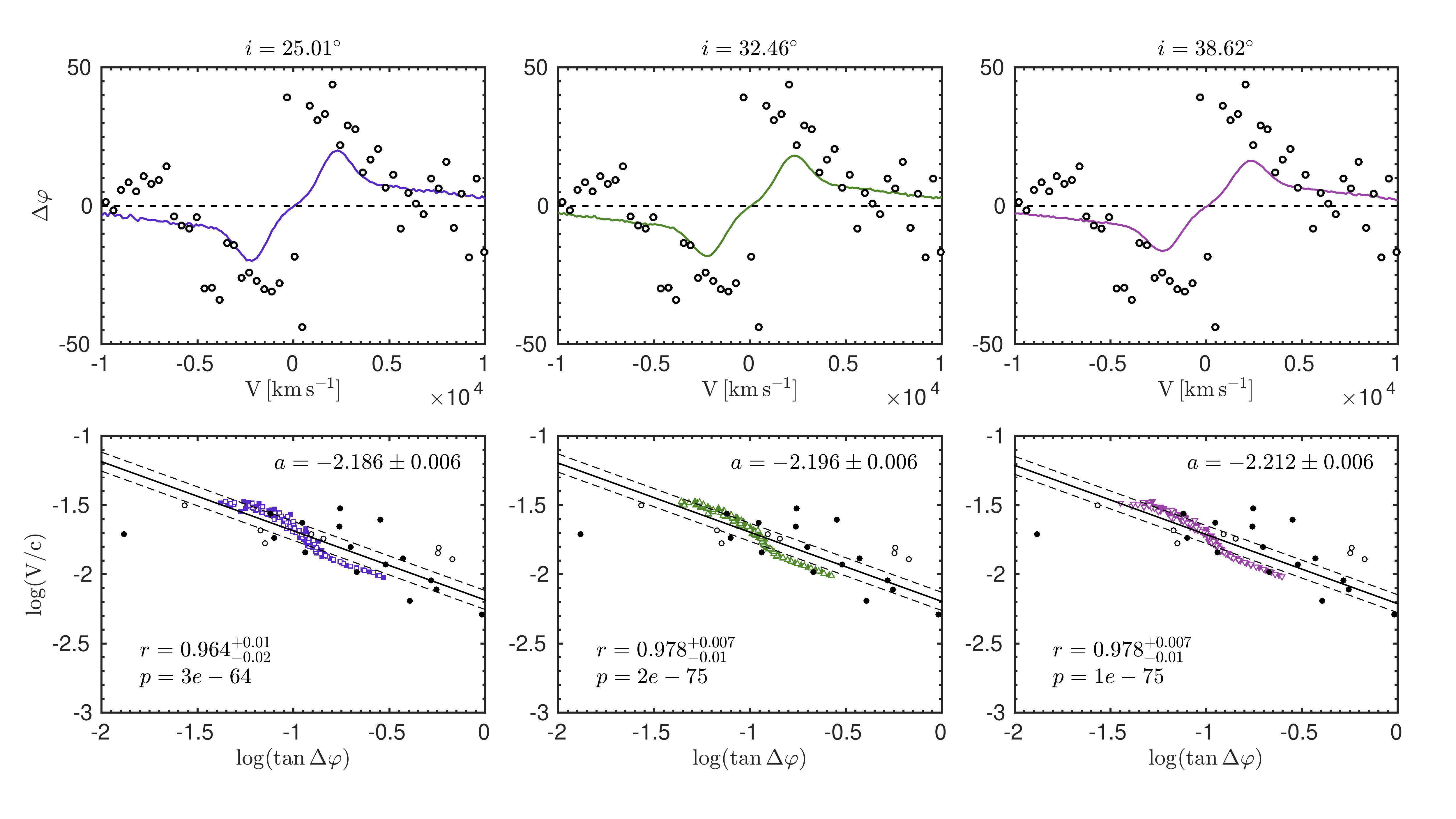}
    \caption{The same as in Fig \ref{NGC4051}, but for 3C273.}
    \label{3C273}%
\end{figure*}

\noindent\textbf{3C 273} -- We obtained very low $\varphi$ dependence on inclination, however, the $\varphi$ amplitude is much smaller, around $19^\circ$ for all inclinations (Fig.\@ \ref{3C273}). Model data show deviation from the straight line in the line wings, however, the scatter is much smaller than it is for the observational data. The ratio $\Rblr{in}{SR}/\Rblr{out}{BLR}\approx 2.19$ is the lowest when compared with other observed objects.

\noindent Mass estimates follow previous trend -- the highest estimate for the lowest inclination.

\begin{figure*}
    \centering
    \includegraphics[width=\hsize]{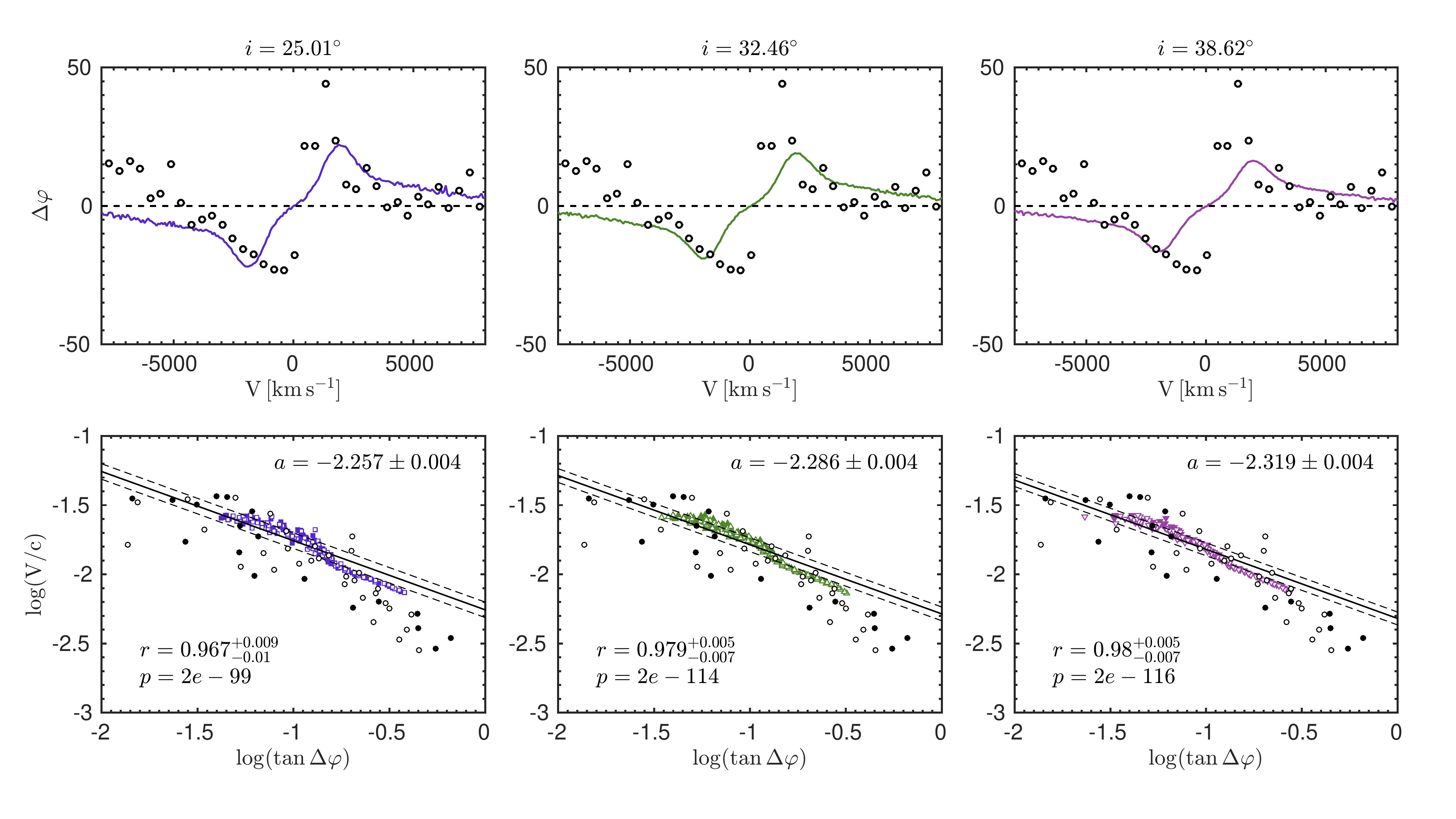}
    \caption{The same as in Fig \ref{NGC4051}, but for PG0844+349.}
    \label{PG0844}%
\end{figure*}

\noindent\textbf{PG0844+349} -- We can see from observations that $\varphi$ profile is asymmetric and that $\varphi$ amplitude is greater for the red part of the line than for the blue part. Results are similar as for the first two objects (Fig.\@ \ref{PG0844}), $\Rblr{in}{SR}/\Rblr{out}{BLR}\approx 2.44$.

For all modeled objects, we were able to produce very similar profiles of $\varphi$ as the observed ones (Figs.\@ \ref{NGC4051}, \ref{NGC4151}, \ref{3C273}, \ref{PG0844}, upper panels). SMBH masses estimated from the fitting of the model data are higher than the ones obtained by fitting the observational data and the obtained values are decreasing as the viewing inclination increases (Table \ref{datamodel}). This is due to the fact that the $\varphi$ amplitude is very sensitive to inclination and is decreasing when viewing from face-on towards edge-on inclinations (from lower to greater). For all observed objects, modeled PO ranges from 0.5\% to 1.5\% for inclinations from lowest to highest. As a measure of the strength of a linear association between the model data and the fit, we give the values of the Pearson correlation coefficient $r$. For all objects we find that the correlation coefficient $r$ is greater than 0.9, except for NGC4151 when viewed from inclination $i = \ang{25.01}$ (Fig.\@ \ref{NGC4151}, lower leftmost panel). The corresponding $p$ values are very close to 0, indicating a strong linear connection between the modeled data and the fit.

Observational data are much more scattered from the predicted straight line. In general, this yields an error in SMBH estimates few times greater compared with an error obtained by reverberation mapping \cite{2015ApJ...800L..35A}. In the case for NGC 4051 and NGC 4151, modeled data fits the best for the highest inclination (Figs.\@ \ref{NGC4051}, \ref{NGC4151}, rightmost bottom panels), or is in offset when viewing more towards face on (Figs.\@ \ref{NGC4051}, \ref{NGC4151}, bottom left and middle panels). The $1\sigma$ uncertainty is smaller for intermediate inclinations. The largest overestimate of the mass, by a factor of 3, is for NGC4051 for $i = \ang{25}$. For all models $\Rblr{in}{SR}/\Rblr{out}{BLR}>2$. This falls in the regime where the SMBH mass estimation shows dependence on inclination. We achieve the best SMBH mass estimates for inclinations $i \approx \ang{39}$ which is close to the value of an average inclination $(i = \ang{39}$ for Type-1 objects \citep{2010ApJS..187..416L,2012MmSAI..83..146H}.

\begin{table*}
\caption{Viewing inclinations, SMBH masses estimates from the model (Column 3), from observations (Column 4) and from reverberation mapping (Column 5).}
\centering
\begin{tabular}{lcccccc}
\hline\hline
Object & $i(^\circ)$ & $\log(\ind{M}{MOD}/\ind{M}{\odot})$ & $\log(\ind{M}{POL}/\ind{M}{\odot})$ & $\log(\ind{M}{REV}/\ind{M}{\odot})$ \\
\hline
            & 25.01 & $7.2 \pm 0.2$\\
NGC 4051    & 32.46 & $6.92 \pm 0.09$ & $6.69 \pm 0.21$ & $6.24 \pm 0.13$\\
            & 38.62 & $6.78 \pm 0.06$\\
\hline
            & 25.01 & $7.56 \pm 0.07$\\
NGC 4151    & 32.46 & $7.40 \pm 0.03$ & $7.21 \pm 0.27$ & $7.12 \pm 0.05$\\
            & 38.62 & $7.27 \pm 0.04$\\
\hline
            & 25.01 & $8.94 \pm 0.09$\\
3C 273      & 32.46 & $8.90 \pm 0.09$ & $8.85 \pm 0.27$ & $8.83 \pm 0.11$\\
            & 38.62 & $8.87 \pm 0.08$\\
\hline
            & 25.01 & $8.00 \pm 0.08$\\
PG0844+349  & 32.46 & $7.95 \pm 0.06$ & $7.70 \pm 0.23$ & $7.85 \pm 0.21$\\
            & 38.62 & $7.88 \pm 0.06$\\
\hline
\end{tabular}
\tablefoot{$\ind{M}{POL}$ denotes masses obtained using AP15 which we used as input mass. Masses obtained by reverberation mapping were taken from \cite{2015PASP..127...67B} using virial factor $\langle f\rangle = 4.31 \pm 1.05$ \citep{2013ApJ...773...90G}.}
\label{datamodel}
\end{table*}

\section{Discussion}
Previous spectropolarimetric studies of Type-1 Seyferts have shown that the polarization signature across the broad $\mathrm{H\alpha}$ is widely varying from object to object \citep{2002MNRAS.335..773S}. At intermediate viewing inclinations, equatorial scattering is dominating the observed polarization and the wavelength averaged polarization $\varphi$ is closely aligned with the projected radio source axis. In their original model, \citet{2005MNRAS.359..846S}, have used single scattering approximation i.e.\ photons emitted from the BLR are being scattered only once from the SR before finally reaching the observer. In their model, SR is optically thin and we find that optical depth of at least 1 along with the higher covering factor of the SR is required to in order to obtain $\varphi$ and PO comparable with observations.

In our Monte Carlo simulations, the treatment of multiple scattering events was fully performed and we find that the largest fraction of photons is being scattered only once, while the other, smaller fraction of photons is undergoing a backward scattering from one side of the SR to the other. This secures good circumstances for the application of the AP15 method. In our model, we approximated the emission of an accretion disk as a point source of isotropic continuum emission. We know that this is not the case and that anisotropy arises due to change in the projected surface area and due to limb darkening effects \citep{1987MNRAS.225...55N}. The strongest emission is in the direction perpendicular to the disk and is rapidly decreasing towards edge-on viewing angles. The inner radius of the SR thus cannot be constant, and should follow similar dependence on the polar angle as the disk emission \citep{2016MNRAS.458.2288S}. Silicate and graphite dust grains have different sublimation temperatures. Graphite grains can survive up to $\sim\SI{1900}{\kelvin}$ and therefore reach closer than silicates which are destroyed when the temperature is $\sim\SI{1200}{\kelvin}$. Furthermore, smaller dust grains are destroyed at the lower temperatures than the larger grains \citep{1984ApJ...277L..71D,1984ApJ...285...89D,1987ApJ...320..537B}. Therefore, we can expect an entire sublimation zone, from graphite to silicate and from larger to smaller dust grains \citep{2007A&A...476..713K,2012MNRAS.420..526M}. This gives opportunity for dust particles to inhabit equatorial region in the close vicinity of the BLR. Equatorial scattering of broad lines from the adjacent SR gives very low inclination dependence on the parameter $a$ rendering the AP15 method inclination independent.

For SMBH mass estimates using the AP15 method, the inner radius of the torus is needed. It can be obtained directly using dust reverberation in the infrared \citep{2011A&A...536A..78K}. The number of objects for which dust reverberation has been performed is smaller than the number of objects for which the reverberation have been performed in optical. For most of the objects, $\Rblr{in}{SR}$ can be calculated only through scaling relations, which can additionally increase an error in the SMBH estimates. The other way is to calculate $\Rblr{in}{SR}$ from the UV radiation \citep{1987ApJ...320..537B}. For this we need to know a priori the physical and chemical composition of dust. Using the right value is important since estimated BH mass is directly proportional to the inner radius.

Seyfert 1 galaxies are often highly variable and when they are in a state of the minimum activity (up to the Type-2), the shape of the position angle of the polarization plane cannot be detected because of the weakness or absence of a flux from the broad line. In this case, the mass estimation using AP15 method is not applicable, even if the object is confidently assigned to the type of objects with equatorial scattering. Therefore, in future modeling, one must take into account the variability of an AGN.

Originally the AP15 method was proposed for systems with an inclination between $20^\circ$ and $70^\circ$. For high viewing inclinations, we have Type-2 objects for which polar scattering dominates the polarization signal and the methods is not longer valid. For almost pole-on AGN, the AP15 method faces two problems. First the amount of interstellar polarization can dominates over the amount of scattering-induced radiation from the innermost regions of AGN. The amount of interstellar polarization is wavelength-dependent and often maximum in the optical band \citep[see][]{1975ApJ...196..261S}. Since the polarization signal of polar AGN in the optical band is usually much lower than 1\% in the optical \citep{2002MNRAS.335..773S,2014MNRAS.441..551M}, the AP15 method is thus restricted to inclinations higher than 20 degrees. Second, the method overestimates SMBH mass by a factor of 1.5 in comparison with the value obtained for the lowest inclination when $\Rblr{in}{SR}/\Rblr{out}{BLR}\approx2$. When the SR is closer, inclination effect is lower and the mass estimates are only depending on the SR inner radius.

Note here that several recent works \citep{2015MNRAS.454.1157P,2016MNRAS.458L..69B,2018MNRAS.473L...1S} gave some ideas to use the spectropolarimetry to estimate BH mass in AGNs. Basically, all above mentioned papers try to constraint the virial factor \citep{2015MNRAS.454.1157P,2018MNRAS.473L...1S}, or to use the broad polarized line \citep{2016MNRAS.458L..69B} to find the black hole mass. In the work by \citet{2018MNRAS.473L...1S}, the authors performed Monte Carlo simulations for a wide range of parameters assuming a static flared-disk geometry for the equatorial region. In comparison to the unpolarized spectra, the virial factor of the polarized spectra has a much narrower distribution. Besides, the half opening angle of the BLR and the nucleus inclinations appear to be the two parameters with the highest influence on the virial factor. The difference between the methods mentioned above and AP15 is that the AP15 method provide direct measuring of the BH mass from the polarization angle, and here we also used some similar approaches as in \citet{2018MNRAS.473L...1S}, but focusing on a polarization plane position angle $\varphi$ and the limits of the AP15 method. In comparison, our 3D polarized radiative transfer simulations have shown that the polarization plane position angle is largely affected by the distance between the BLR and SR. If the two share similar values, the mass estimated using AP15 method becomes inclination independent, which is a great advantage in comparison to traditional reverberation mapping techniques.

\section{Conclusions}
We modeled polarization effects in AGN broad lines in order to constrain the limits of the AP15 method for the BH estimates using polarization in broad lines.

We used Monte Carlo radiative transfer code \textsc{stokes} that includes multiple scattering for accurate polarization treatment. We considered equatorial scattering (on the torus) of the light from a BLR that has dominant Keplerian motion. Additionally we considered complex BLR kinematics having inflows and outflows

We explore all these effects on the accuracy of the BH mass measurement using AP15 method.

From our investigation we can outline following conclusions:

\begin{itemize}
 \item If Keplerian motion can be traced through the polarized line profile, then direct estimates of the mass can be performed for obtaining reasonable values.
 \item The effects of possible inflow/outflow configuration of the BLR take its toll only for extreme cases where the velocity of inflowing/outflowing emitter is comparable or higher than the Keplerian velocity.
 \item Masses of the SMBHs obtained by AP15 method are in a good agreement with the ones found in literature.
\end{itemize}

The AP15 method gives us the new independent way of mass estimation. Future parameter grid will be extended with inflow/outflow configuration of the scattering region with the possibility of considering clumpy structures. We expect to perform high quality spectropolarimetric observations of the high-redshifted quasars and test the AP15 method on high ionized lines such as C III] and C IV.

\begin{acknowledgements}
      \textbf{We thank an anonymous referee for constructive suggestions that improved this paper.} This work was supported by the Ministry of Education and Science (Republic of Serbia) through the project Astrophysical Spectroscopy of Extragalactic Objects (176001), the French PNHE and the grant ANR-11-JS56-013-01 “POLIOPTIX” and the Russian Foundation for Basic Research grant N15-02-02101 and N14-22-03006. D.\@ Savi\'c thanks the French Government and the French Embassy in Serbia for supporting his research. Part of this work was supported by the COST Action MP1104 “Polarization as a tool to study the Solar System and beyond”.
\end{acknowledgements}


%
%

\bibliographystyle{aa} 

\end{document}